\documentclass[aps, prb, twocolumn, superscriptaddress, showpacs]{revtex4}
\usepackage[german,american,english]{babel}
\usepackage{graphicx}
\usepackage{graphics}
\usepackage{dcolumn}
\usepackage{bm}
\usepackage{amssymb}
\usepackage{amsmath}
\usepackage{amsfonts}
\usepackage{epsfig}
\usepackage{braket}
\usepackage{url}
\usepackage{multirow}
\usepackage{color}

\begin{document}

\title{Spin-valley qubit in nanostructures of monolayer semiconductors: Optical control and hyperfine interaction}

\author{Yue Wu}
\affiliation{Department of Physics and Center of Theoretical and Computational
Physics, The University of Hong Kong, Hong Kong, China}
\author{Qingjun Tong}
\affiliation{Department of Physics and Center of Theoretical and Computational
Physics, The University of Hong Kong, Hong Kong, China}
\author{Gui-Bin Liu}
\affiliation{School of Physics, Beijing Institute of Technology, Beijing 100081, China}
\author{Hongyi Yu}
\affiliation{Department of Physics and Center of Theoretical and Computational
Physics, The University of Hong Kong, Hong Kong, China}
\author{Wang Yao}
\affiliation{Department of Physics and Center of Theoretical and Computational
Physics, The University of Hong Kong, Hong Kong, China}

\begin{abstract}
We investigate the optical control possibilities of spin-valley qubit carried by single electrons localized in nanostructures of monolayer TMDs, including small quantum dots formed by lateral heterojunction and charged impurities. The quantum controls are discussed when the confinement induces valley hybridization and when the valley hybridization is absent. We show that the bulk valley and spin optical selection rules can be inherited in different forms in the two scenarios, both of which allow the definition of spin-valley qubit with desired optical controllability. We also investigate nuclear spin induced decoherence and quantum control of electron-nuclear spin entanglement via intervalley terms of the hyperfine interaction. Optically controlled two-qubit operations in a single quantum dot are discussed.
\end{abstract}

\maketitle

\section{Introduction}
\label{sec:Intro}

Single spins at semiconductor nanostructures have been widely explored as information carriers in quantum computing, quantum spintronics and quantum metrology. \cite{awschalom2007diamond, kane1998silicon, jelezko2006single}
These solid state qubit systems of interest include the spin of single electrons or holes localized at quantum dots or by impurities formed in various bulk semiconductors such as the III-V compounds, silicon and diamond, and their heterostructures and nanoscrytals. These electron and hole spin qubits have demonstrated remarkable optical and electrical controllability, relatively long coherence time at low temperature compared to the unit operation time, and potential integrability with existing semiconductor technologies. Through the hyperfine interactions with the electron or hole spin qubits, the lattice nuclear spins also play crucial roles in these solid state qubit systems, either as additional information carriers with the advantage of extremely long storage time, or as a deleterious noise source that need to be suppressed.

Atomically thin two-dimensional (2D) semiconductors offer new opportunities for quantum spintronics and spin based quantum information processing. An electrically controllable spin qubit system based on 2D materials was first proposed in graphene, a gapless semiconductor. \cite{trauzettel2007spin}
Monolayer group-VIB transition metal dichalcogenides (TMDs) have recently emerged as a new class of direct gap 2D semiconductors with appealing optical properties and rich spin physics, implying their great potentials for hosting optically controlled spin qubits. \cite{wang2012electronics, xu2014spin}
These compounds are of the chemical composition of MX$_2$ (M = Mo,W; X = S, Se). The monolayer is a X-M-X covalently bonded hexagonal 2D lattice, with a direct bandgap in the visible frequency range which is ideal for optoelectronic applications and for the exploration of optical control of spin. \cite{splendiani2010emerging, mak2010atomically}
Single electrons can be confined in quantum dots defined by lateral confinement potentials on an extended monolayer, e.g. by patterned electrodes, similarly to the quantum dots in III-V heterostructures, and electrically controlled spin qubit has been proposed. \cite{hanson2007publisher, kormanyos2014spin}
Alternatively, quantum dot confinement can also be realized by the lateral heterojunctions on a single crystalline monolayer, but with different metal elements in and outside the quantum dot region (c.f. Fig. 1), where the band offset between the different TMD compounds forms the potential to confine single electron or hole. Lateral heterostructures with MoSe$_2$ islands surrounded by WSe$_2$ on a crystalline monolayer have been demonstrated very recently using chemical vapor deposition growth, although the length scale of the island is $\sim \mu$m, still too large for confining single electron. \cite{huang2014lateral}
The TMDs monolayers also host various atomic defects which can localize single electron or hole as well. \cite{zhou2013intrinsic, feng2014effect, fuhr2004scanning, noh2014stability, tongay2013defects}
Remarkably, recent experiments have shown that certain types of defects in monolayer WSe$_2$ are excellent single photon sources, emitting at an energy which is a few tens meV below the exciton in the 2D bulk. \cite{srivastava2015optically, he2015single, koperski2015single, chakraborty2015voltage, Tonndorf15single}
Such behaviors of the TMDs defects resemble the shallow impurities in conventional semiconductors (e.g. Si donor in III-V compounds) that localize single electron (or hole) as well as single exciton, implying the possibility towards optical control of single electron spin. \cite{Yamamoto09}

Optically controlled spin qubit is highly appealing in monolayer TMDs because of the interesting optical properties of the 2D bulk. The monolayer TMDs have a unique band structure with the conduction and valence band edges both at the degenerate K and -K valleys at the corners of the hexagonal Brillouin zone. The direct-gap optical transitions have a selection rule: left- (right-) handed circular polarized photon couples to the interband transitions in the K (-K) valley only. \cite{Yao08PRB77,xiao2012coupled}
This valley dependent optical selection rule has made possible in the 2D bulk the optical pumping of valley polarization, \cite{Zeng2012valley, TonyHeinzNanotech7, Cao2012valley}
and optical generation of valley coherence. \cite{Jones2013optical}
Moreover,  the spin-orbit coupling from the metal atoms gives rise to a pronounced coupling between the valley pseudospin and spin, \cite{Gong2013magneto, Jones2014spin, xu2014spin}
through which the optical selection rule becomes a spin dependent one, allowing the optical control of spin as well. This suggests that the valley pseudospin and spin of a single electron can be a promising qubit carrier with optical controllabilities, as long as these bulk properties can be inherited when the electron is localized in the monolayers.

A major difference in the spin and valley pseudospin physics expected between the bulk electron and the localized electron is the intervalley coupling and valley hybridization by the confinement. This issue has been studied for quantum dot confinement potentials on extended monolayers, \cite{liu2014intervalley}
where the intervalley coupling is found to be weak for quantum dot with lateral size larger than 20 nm ($\sim$ meV or orders smaller, depending on the shape and size of the dot). In such a case, the valley hybridization is well quenched by the much stronger spin-valley coupling, and the quantum dot can well inherit the valley and spin physics of the 2D bulk. Interestingly, a sensitive dependence of intervalley coupling strength on the central position of the confinement potentials is discovered. It is found that when the potential has C$_3$ or higher rotational symmetry, the intervalley coupling completely vanishes if the potential center is at a chalcogen atom site or the hollow center of the hexagon formed by metal and chalcogen atoms, which is due to the dependence of the eigenvalue of C$_3$ rotation operator on the location of the rotation center. \cite{liu2014intervalley, liu2015electronic}

In this work, we investigate the optical control possibilities of spin-valley qubit carried by single electrons localized in nanostructures of monolayer TMDs, including charged impurities and small quantum dots (with length scale of 10 nm or smaller). We discuss the quantum controls under two different scenarios: (i) in the presence of valley hybridization due to the strong confinement and (ii) in the absence of valley hybridization. The latter scenario is considered for the confinements that has C$_3$ or higher rotational symmetry about a chalcogen atom site or the hollow center of the hexagon formed by metal and chalcogen atoms, or when this symmetry is only weakly broken so that the residue intervalley coupling can be well quenched by the spin-valley coupling. We show that the bulk valley and spin optical selection rules can be inherited in different forms in the two scenarios, both of which allow the coherent rotation between electron states controlled by optical pulses. The hyperfine interaction between lattice nuclear spins and the electron or hole spin is also formulated within the envelop function approximation, and the nuclear spin induced decoherence of the spin-valley qubit is analyzed. We find that the short-range nature of the hyperfine interaction gives rise to intervalley terms, which can be utilized for optical control of the electron-nuclear spin entanglement.

The rest of the paper is organized as follows. In Sec. \ref{nanoconfine}, we give a brief account of the nanostructures being considered here for optically controlled spin-valley qubit. In Sec. \ref{withvalleyhybridization}, we discuss the electron states in presence of valley hybridization expected in strong confinement, and present the optical selection rules for the quantum confined states. Coherent rotations between valley hybridized states by optical control will be discussed. In Sec. \ref{withoutvalleyhybridization}, we discuss the electron states in the absence of valley hybridization when the confinement has the aforementioned rotational symmetry. The optical control is achieved with the help of external magnetic fields. In Sec. \ref{hyperfine}, we discuss the hyperfine interactions of the confined electrons and holes with lattice nuclear spins in the envelope function approximation. We propose to optically control the electron-nuclear spin entanglement via intervalley terms of the hyperfine interaction. The decoherence time of the localized electron spin caused by interacting with lattice nuclear spins is discussed. In Sec. \ref{addsection}, we discuss the possibility to realize two-qubit logic operations between the spin qubit and the valley qubit carried by a single electron in a quantum dot. Finally, conclusions are given in Sec. \ref{Conclusions}. Appendix \ref{intervalleycoupling} uses a three-band tight-binding model to estimate the intervalley coupling strength in the confinement by charged impurity and small quantum dot. In Appendix \ref{hyperfinedetail}, we analyze the terms in the electron-nuclear and hole-nuclear hyperfine interactions based on symmetries of the relevant atomic orbitals, and estimate the bulk hyperfine constants.

\section{Confinement of single electron in the nanostructures\label{nanoconfine}}

If the length scale of the confinement potential is still much larger than the lattice constants, the bound states are formed predominantly from the band-edge Bloch states in the K and -K valleys of the 2D bulk. In general, each eigenstate in the confinement is a hybridization of Bloch states from the K and -K valleys due to the intervalley coupling introduced by the confinement potential, except when the potential has a $C_3$ rotational symmetry about either a chalcogen atom site or a hollow center of the hexagon formed by metal and chalcogen atoms.\cite{liu2014intervalley} In the absence of valley hybridization, the bound-state eigenfunctions can be constructed from Bloch states from the K or -K valley only, denoted as $\Psi_{\tau,s}$ where $\tau = \pm$ is the valley index for the $\pm$K valley, and $s = \uparrow$ ($\downarrow$) denotes spin up (down) state. This is a convenient basis for our discussion, even when intervalley coupling and valley hybridization are present. Intervalley coupling is then the off-diagonal matrix elements between these basis states due to the confinement potential. If confinement potential is spin-independent, intervalley coupling vanishes between states with opposite spin index.
With the large quantization energy in the confinement potential (see Appendix \ref{intervalleycoupling}), we can focus only on the ground states for each spin and valley index, while the excited states are far off resonance concerning either the valley hybridization effect or the optical control of the spin-valley qubit. Below we consider two types of confinements.
\begin{figure*}[tbp]
\includegraphics[width=13 cm]{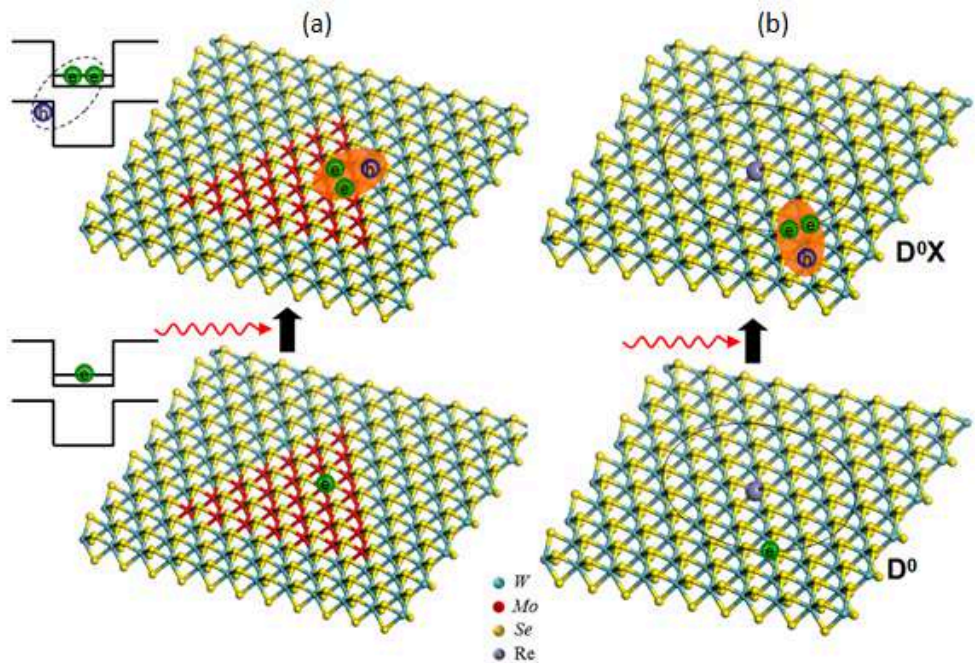}
\caption{Schematics of nanostructures of monolayer TMDs for optically controlled spin-valley qubit: (a) small quantum dots in lateral heterojunction and (b) charged impurity system. In (a), the heterojuction is formed of a MoSe$_2$ island in monolayer WSe$_2$. An electron is confined in the MoSe$_2$ quantum dot (bottom panel) and is excited to the trion state (top panel). The schematics of the confinement potential is shown on the left. In (b), the impurity system is formed in monolayer WSe$_2$ when a W atom is replaced by a Re one. The $D^0$ system is shown at the bottom and $D^0X$ is shown at the top.}
\label{Trion}
\end{figure*}

The first is the lateral heterostructures between different TMDs, for example, a MoSe$_2$ island surrounded by WSe$_2$ on a crystalline monolayer. With a type-II band alignment between the two TMDs, such a heterostructure forms a confinement potential of a vertical wall, which localizes a single electron in the MoSe$_2$ region with a potential depth of few hundred meV. The valley and spin degrees of this electron can then define a qubit. In the optical control, an optical field can couple the states of the single electron to the optical excited states of trion (i.e. two electrons plus a hole) through the interband transition. These trion states are utilized as intermediate states for the optical control of the single electron states. Although the heterostructure itself does not form a confinement for a single hole (c.f. Fig. \ref{Trion}), the quantum confinement of the electron constituents will nevertheless localize the trion at the heterostructure.

The intervalley coupling strength grows with the decrease in size of the quantum dot. At a lateral size of 5 nm, the coupling matrix element reaches $0.1-1$ meV depending on the quantum dot shape. The valley hybridization will then be determined by the competition of this off-diagonal matrix elements in the basis $\Psi_{\tau,s}$, and the diagonal energy differences between $\Psi_{+,s}$ and $\Psi_{-,s}$ due to the spin-valley coupling in the  band of the 2D bulk. For monolayer MoS$_2$, the spin-valley coupling strength in the conduction band is 3 meV, \cite{liu2013three}
comparable to the achievable intervalley coupling in small quantum dots. For other three TMDs (MoSe$_2$, MoS$_2$ and WSe$_2$), the spin-valley coupling strength is in the range of 20-40 meV. Valley hybridization for the hole component is always negligible due to the giant spin-valley coupling of hundred of meV in all TMDs.

The second type of nanostructure is a neutral donor system $D^0$, for example a Re replacing a W in the WSe$_2$ monolayer, where the positively charged impurity binds the extra electron and forms a hydrogenic state. Similarly to the quantum dot case, the single electron states can be optically coupled to the donor bound exciton $D^0X$ states. In GaAs, $D^0$-$D^0X$ system has been extensively explored for optically controlled single spin.\cite{Yamamoto09} Compared with the quantum dot, the $D^0$-$D^0X$ system in monolayer TMDs is expected to be a much tighter confinement due to the enhanced Coulomb interaction. Consequently, the intervalley coupling strength is much stronger (unless the impurity is centered at a chalcogen atom site or a hollow center of the hexagon formed by metal and chalcogen atoms). For several examplary electrostatic Coulomb potential as shown in Appendix \ref{intervalleycoupling}, we find the intervalley coupling strength can be comparable to the electron spin-valley coupling strength in MoSe$_2$, MoS$_2$ and WSe$_2$. Therefore, the valley hybridization of electron is expected to be significant in the $D^0$-$D^0X$ system.

\section{Optical control of electron states in presence of valley hybridization \label{withvalleyhybridization}}

\begin{figure}[tbp]
\centering
\includegraphics[width=3.3 in,height=2.4 in]{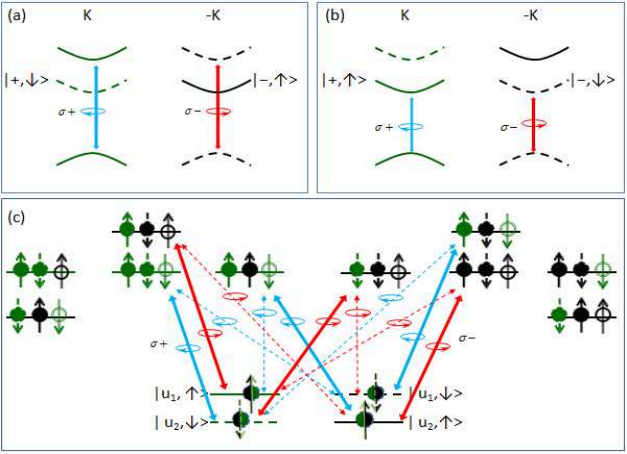}
\caption{Valley and spin dependent optical transition selection rules at the band edge of the monolayer WX$_2$ (a) and MoX$_2$ (b). (c) Optical transition selection rules in quantum dots with valley hybridization of the localized single electron. Solid (dashed) horizontal lines denote spin-up (-down) states. Dark green (black) color denotes $+K$ ($-K$) valley. Red doubled-arrowed lines denote $\sigma^-$-polarized lights and blue ones denote $\sigma^+$-polarized lights, with the transition strength in dashed one $\propto \sin \frac{\theta}{2}$ (i.e. enabled by a finite intervalley coupling, see text), and solid one $\propto \cos \frac{\theta}{2}$.} \label{fig2}
\end{figure}
In this section, we consider the scenario where the confinement potential introduces pronounced valley hybridization of the localized electron. This applies to the confinement potentials of small length scale which do not have the $C_3$ rotational symmetry about either a chalcogen atom site or a hollow center of the hexagon formed by metal and chalcogen atoms (see Sec. ~\ref{nanoconfine} and Appendix \ref{intervalleycoupling}).

We note that valley hybridization is present for electrons only. For holes, the band edges of the 2D bulk are spin-valley locked because of the giant spin-orbit coupling, i.e. valley K (-K) has spin down (up) holes only. As the confinement potential does not flip spin, valley hybridization by the confinement is completely quenched for the spin-valley locked holes.  For electrons with a much smaller spin-valley coupling in the 2D bulk band edges, we take into account both spin species in each valley, and the quantum dot Hamiltonian in the aforementioned basis is,
\begin{equation}
H_0=h \tau_x+\frac{\lambda}{2}\tau_z s_z \label{t1},
\end{equation}
where $h$ is the intervalley coupling strength, $\tau$ and $s$ denote the pauli matrices operating at valley and real spin degrees of freedom, and $\lambda$ is spin-valley coupling of conduction band.

Since intervalley coupling conserves spin, we re-write the Hamiltonian in a compact form,
\begin{eqnarray}
H_0&=&\vec{d} \cdot \vec{\tau}=d \left(
                                                 \begin{array}{cc}
                                                   \cos\theta & \sin\theta \\
                                                   \sin\theta & -\cos\theta \\
                                                 \end{array}
                                               \right)
 \label{Hvector}
\end{eqnarray}
with $\vec{d}=(h,0,\frac{\lambda}{2}s)= d (\sin\theta,0,\cos\theta)$. The eigenenergies are $\epsilon_{1(2),s} =\pm s d $, with eigenvectors,
\begin{equation}
\left\vert u_1,s\right\rangle=\left(
\begin{array}{cc}
\cos\frac{\theta}{2} \\
\sin\frac{\theta}{2}
\end{array}%
\right), \label{u1}
\end{equation}
\begin{equation}
\left\vert u_2,s\right\rangle=\left(
\begin{array}{cc}
\sin\frac{\theta}{2} \\
-\cos\frac{\theta}{2}
\end{array}%
\right). \label{u2}
\end{equation}
These four spin-valley configurations of the single electron can be used to construct the qubit.

Our proposed optically controlled qubit operations rely on the optical selection rules in monolayer TMDs. \cite{xiao2012coupled}
In 2D bulk of monolayer TMDs, the conduction (valence) band edge states mainly consist of transition metal $d_{z^2}$ ($d_{x^2-y^2}\pm i d_{xy}$) orbitals with the magnetic quantum $m_c=0$ ($m_v=\pm2$). At the $\pm K$ points, the Bloch states have $C_3$ rotation symmetry $C_3\left\vert \tau,s\right\rangle=e^{-i\frac{2m\pi}{3}}\left\vert \tau,s\right\rangle$, which implies an azimuthal selection rule for the allowed interband optical transition ($m_c-m_v\mp1$ modulo $3$)$=0$. Because of inversion symmetry breaking, this optical selection rules is valley-contrasted. The spin-valley locking of the holes further makes these selection rules spin-dependent: $\sigma^+$ circular polarization optical field can generate spin-up electrons and spin-down holes in valley $K$, while the excitation in the $-K$ valley is simply the time-reversal of the above, as shown in Fig. \ref{fig2} (a) for WX$_2$ systems and Fig. \ref{fig2} (b) for MoX$_2$ systems. Since these two kinds of systems only differed by the sign of spin
splitting in the conduction band \cite{liu2013three}, we illustrate our results with WX$_2$ system in all of the following figures.

In the context of a quantum dot charged with a single electron, an optical field can couple the different spin-valley states of the single electron to an charged exciton state (trion) of the various spin-valley configurations. These transitions have optical polarization selection rules inherited from the 2D bulk. With the valley hybridization of electrons, the optical transitions in fact become more intricate in the quantum dot. As shown in Fig. \ref{fig2} (c), there are six bright trion states that can be coupled to the four spin-valley states of the single electron. The dashed arrows denote the transitions with strength $\propto \sin \frac{\theta}{2}$ (i.e. enabled by a finite intervalley coupling $h$), while the solid arrows denote the transitions with strength $\propto \cos \frac{\theta}{2}$.

\begin{figure}[tbp]
\centering
\includegraphics[width=3.3 in,height=1.2 in]{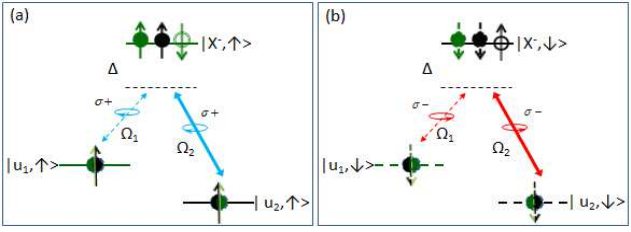}
\caption{Optically controlled rotation between the valley hybridized states via the Raman type processes mediated by the trion state. There are two $\Lambda$-type three-level systems that can be controlled respectively via the Raman type process by optical pulses with $\sigma^+$ polarization (a) or $\sigma^-$ polarization (b). The quantum dot can have four states for encoding information that are distinguished by the spin index and can be selectively accessed using circularly polarized light.} \label{fig3}
\end{figure}
Among all the possible optical transitions, we note that $\left\vert u_1,\uparrow\right\rangle$ and $\left\vert u_2,\uparrow\right\rangle$ can both be coupled to the same trion state $\left\vert X^-,\uparrow \right\rangle=e^{\dag}_{+,\uparrow}e^{\dag}_{-,\uparrow}h^{\dag}_{+,\Downarrow}\left\vert G \right\rangle$ with a $\sigma^+$ circular polarized light. Here $e^{\dag}_{\tau,s}$ creates an electron state with spin $s$ and valley $\tau$ and similarly $h^{\dag}_{\tau,s'}$ creates a hole state with spin $s'=\Uparrow,\Downarrow$ with $\left\vert G\right\rangle$ denoting an empty conduction band and full valence band. Therefore $\left\vert u_1,\uparrow\right\rangle$, $\left\vert u_2,\uparrow\right\rangle$ and $\left\vert X^-,\uparrow\right\rangle$ form a $\Lambda$-type three-level system (c.f. Fig. \ref{fig3} (a)). Similarly, a $\sigma^-$ circular polarized light couples $\left\vert u_1,\downarrow\right\rangle$ and $\left\vert u_2,\downarrow\right\rangle$ with $\left\vert X^-,\downarrow\right\rangle=e^{\dag}_{+,\downarrow}e^{\dag}_{-,\downarrow}h^{\dag}_{-,\Uparrow}\left\vert G \right\rangle$, forming another $\Lambda$-type three-level system (c.f. Fig. \ref{fig3} (b)). We note that a single quantum dot can now have four states for encoding information: \{$\left\vert u_1,\uparrow\right\rangle$, $\left\vert u_2,\uparrow\right\rangle$, $\left\vert u_1,\downarrow\right\rangle$, $\left\vert u_2,\downarrow\right\rangle$\}. $\sigma^+$ or $\sigma^-$ polarized light makes possible selective access of this Hilbert space for either initialization, readout, or quantum control, where optical control scenarios utilizing the $\Lambda$ level scheme can be borrowed from optically controllable III-V quantum dots.\cite{economou2006proposal,liu2010quantum}

For example, coherent rotation between the pair of states \{$\left\vert u_1,\uparrow\right\rangle$, $\left\vert u_2,\uparrow\right\rangle$\} (or \{$\left\vert u_1,\downarrow\right\rangle$, $\left\vert u_2,\downarrow\right\rangle$\}) can be realized through an optical Raman process via the intermediate trion states $\left\vert X^-,\uparrow\right\rangle$ (or $\left\vert X^-,\downarrow\right\rangle$) by $\sigma^+$ (or $\sigma^-$) polarized light in the $\Lambda$-type three-level system. \cite{chen2004theory} Applying two phase-locked optical pulses with $\sigma^+$ polarization, the light-matter interaction Hamiltonian in the rotating wave approximation is
\begin{equation}
H_I=\sum_{j=1,2}\Omega_j(t)\left\vert X^-,\uparrow\right\rangle\left\langle u_j,\uparrow\right\vert+\text{H.c.},
\end{equation}
where the Rabi frequencies are of the forms $\Omega_{1}(t)=E_{1} D_0 \sin\frac{\theta}{2}e^{i\omega_1t-i\alpha_1}$ and $\Omega_{2}(t)=E_{2} D_0 \cos\frac{\theta}{2}e^{i\omega_2t-i\alpha_2}$ with $E_j$ being the amplitude of the polarized light and $\alpha_1-\alpha_2\equiv\alpha$ being the relative phase between them. $D_0=\left\langle u_j,\uparrow\right\vert D \left\vert X^-,\uparrow\right\rangle$ is the optical transition matrix element between the localized electron state and the trion state, which is approximately proportional to $\frac{\left\langle \tau,s\right\vert D \left\vert \tau,s' \right\rangle}{a_H}$, where $\left\langle \tau,s\right\vert D \left\vert \tau,s' \right\rangle$ is the optical transition matrix element between the bulk conduction and valence states at $\tau K$ points and $a_H$ is the Bohr radius of the trion state. Because of the strong Coulomb interaction,  $D_0$ is several times larger than the one in III-V semiconductor quantum dots.  The frequencies $\omega_j=E_t-\Delta-\epsilon_j$ are chosen to satisfy the Raman conditions with $E_t$ and $\Delta$ being the trion energy and Raman detuning respectively. In the rotating frame defined by $e^{-i\epsilon_1t}\left\vert u_{1},\uparrow\right\rangle$, $e^{-i\epsilon_2t}\left\vert u_{2},\uparrow\right\rangle$ and $e^{-i(E_T-\Delta)t}\left\vert X^-,\uparrow\right\rangle$, the total Hamiltonian $H=H_0+H_I$ is transformed to
\begin{equation}
H=\left(
    \begin{array}{ccc}
      0 & 0 & E_{1} D_0 \sin\frac{\theta}{2}e^{-i\alpha} \\
      0 & 0 & E_{2} D_0 \cos\frac{\theta}{2} \\
      E_{1} D_0 \sin\frac{\theta}{2}e^{i\alpha} & E_{2} D_0 \cos\frac{\theta}{2} & \Delta \\
    \end{array}
  \right) \label{Hcouple}
,
\end{equation}
where the fast oscillating terms $\propto e^{-2idt}$ have been neglected. For large detuning, the trion state is eliminated via using the adiabatic approximation. The dynamics of the qubit is then described by
\begin{equation}
H_{eff}=\frac{-D_0^2}{\Delta}\left(
    \begin{array}{cc}
      E_{1}^2 \sin^2\frac{\theta}{2} & \frac{E_{1}E_{2}}{2} \sin\theta e^{i\alpha} \\
      \frac{E_{1}E_{2}}{2} \sin\theta e^{-i\alpha} & E_{2}^2 \cos^2\frac{\theta}{2} \\
    \end{array}
  \right)
\end{equation}
which can be rewritten as
\begin{equation}
H_{eff}=n_0 I+\vec{n}\cdot\vec{\zeta}, \label{field}
\end{equation}
with
\begin{eqnarray}
n_0&=&-\frac{D_0^2(E_{1}^2 \sin^2\frac{\theta}{2}+E_{2}^2 \cos^2\frac{\theta}{2})}{2\Delta}, \notag \\ \notag
n_x&=&-\frac{E_{1}E_{2} D_0^2 \sin\theta \cos\alpha}{2\Delta}, \\ \notag
n_y&=&\frac{E_{1}E_{2} D_0^2 \sin\theta \sin\alpha}{2\Delta}, \\
n_z&=&-\frac{D_0^2(E_{1}^2 \sin^2\frac{\theta}{2}-E_{2}^2 \cos^2\frac{\theta}{2})}{2\Delta}, \label{n}
\end{eqnarray}
where $\vec{\zeta}$ operates on our defined qubit, which precedes under this pseudo-magnetic field $\vec{n}$.

The effect of intervalley coupling is involved in the angle $\theta$. Without intervalley coupling, $\theta=0$, $\vec{n}$ only lies in $z$ direction. Therefore, intervalley coupling plays an crucial role in the optically controlled single-qubit operation. In general cases with finite intervalley coupling, arbitrary pseudo-magnetic field orientation can be obtained by changing the control parameters $E_{1,2}$, $\alpha$ and $\Delta$. For example, when one of the two pulses is turned off, i.e. $E_1$ or $E_2$ being set to zero, $\vec{n}$ is in $z$ direction. This realizes a single-qubit phase-shift gate $U_{S_{\phi}}=\left(
                                                                                                                                     \begin{array}{cc}
                                                                                                                                       1 & 0 \\
                                                                                                                                       0 & e^{i\phi} \\
                                                                                                                                     \end{array}
                                                                                                                                   \right)$ if we set $E_1=0$, where $\phi=\frac{D_0^2E_2^2\cos^2\frac{\theta}{2}t}{\Delta}$.
On the other hand, when $\frac{E_2}{E_1}=\tan\frac{\theta}{2}$ and $\alpha=0$, $\vec{n}$ is in the $x$ direction with $n_x=-\frac{E_1^2 D_0^2 \sin^2\frac{\theta}{2}}{\Delta}$. The qubit state would be flipped by an optical pulse with duration $t_f=\frac{\pi}{2|n_x|}$. For square-shaped MoS$_2$ quantum dot with lateral size of 3 nm,  the intervalley coupling is calculated as 1 meV if the lateral confinement potential is set as 0.2 eV (see Appendix A). When a light with $E_1 D_0=0.5$ meV is applied, $|n_x|\sim1$ $\mu$eV and $t_f$ is about 0.8 ns, if we set the detuning $\Delta= 5$ meV.

\section{electron states in absence of valley hybridization \label{withoutvalleyhybridization}}

\begin{figure}[tbp]
\includegraphics[width=3.1 in,height=2.5 in]{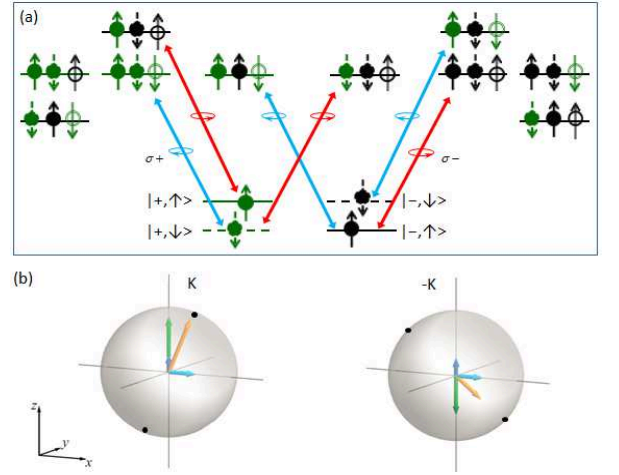}
\caption{(a) Optical transition selection rules in quantum dots without valley hybridization, and in absence of magnetic field. (b) Configurations of effective magnetic fields in the two valleys. The dark blue, light blue, green and yellow arrows indicate $B_z$, $B_x$, $\lambda$ and total field $\vec{d'}$ respectively. In each valley, the eigenstates of the effective magnetic field are indicated (black spots).}
\label{fig4}
\end{figure}

If the confinement has $C_3$ symmetry, intervalley coupling vanishes when the confinement center is chosen at the chalcogen atom site or the hollow center of the hexagon lattice.\cite{liu2014intervalley,liu2015electronic} In this case, valley is a good quantum number, and the quantum dot states are formed from the Bloch states in a single valley of the 2D bulk. The optical transitions of the spin-valley states of the single electrons to trions in Fig. \ref{fig2} (c) then reduces to those in Fig. \ref{fig4} (a).

Optical control of the spin states is still possible in the presence of a magnetic field with an in-plane component, which can couple the spin up and down states from the same valley.
With external magnetic fields, the Hamiltonian for the single electron at each valley becomes
\begin{eqnarray}
H'_0&=&\frac{\lambda}{2}\tau_z s_z +B_x s_x+B_z s_z
=\vec{d'} \cdot \vec{s},
 \label{Hwt}
\end{eqnarray}
where $\vec{d'}=(B_x,0,\frac{\lambda}{2}\tau+B_z)=d'(\sin\theta',0,\cos\theta')$ is the effective field on the spin doublet at each valley, as plotted schematically in Fig. \ref{fig4} (b), which is valley-dependent in general. The eigenstates of this Hamiltonian are
\begin{equation}
\left\vert u'_1,\tau\right\rangle=\left(
\begin{array}{cc}
\cos\frac{\theta'}{2} \\
\sin\frac{\theta'}{2}
\end{array}%
\right), \label{u'1}
\end{equation}
\begin{equation}
\left\vert u'_2,\tau\right\rangle=\left(
\begin{array}{cc}
\sin\frac{\theta'}{2} \\
-\cos\frac{\theta'}{2}
\end{array}%
\right). \label{u'2}
\end{equation}
These spin-coupled states can be used to construct the qubits. In contrast to the scenario in Sec. \ref{withvalleyhybridization} in presence of the valley hybridization, the valley index is now a good quantum number while the spin is now quantized along a direction tilted from z. Coherent rotation between the pair of states \{$\left\vert u'_1,+\right\rangle$, $\left\vert u'_2,+\right\rangle$\} (or \{$\left\vert u'_1,-\right\rangle$, $\left\vert u'_2,-\right\rangle$\}) can be realized through an optical Raman process via the intermediate trion states (c.f. Fig. 5).

\begin{figure}[tbp]
\centering
\includegraphics[width=3.3 in,height=2.3 in]{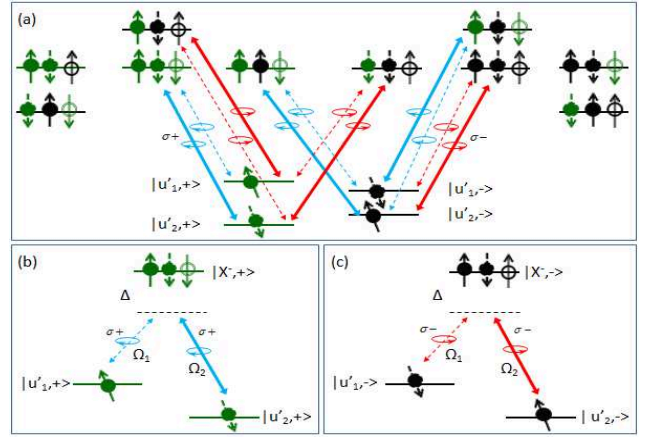}
\caption{(a) Optical transition selection rules in quantum dots without intervalley coupling while with applied magnetic fields. The coupling strength in dashed line $\propto \sin \frac{\theta'}{2}$ and solid one $\propto \cos \frac{\theta'}{2}$. There are two $\Lambda$-type three-level systems that can be controlled respectively via the Raman type process by optical pulses with $\sigma^+$ polarization (b) or $\sigma^-$ polarization (c). The quantum dot can therefore provide four states that are distinguished by the valley index and can be accessed using circularly polarized light.} \label{fig5}
\end{figure}

In an applied magnetic field with a finite in-plane component, there are six bright trion states, as shown in Fig. \ref{fig5} (a). One can see that the two states $\left\vert u'_1,+\right\rangle$ and $\left\vert u'_2,+\right\rangle$ are coupled to the trion state $\left\vert X^-,+\right\rangle=e^{\dag}_{+,\downarrow}e^{\dag}_{+,\uparrow}h^{\dag}_{+,\Downarrow}\left\vert G\right\rangle$ by a $\sigma^+$ polarized light. Similarly, $\left\vert u'_1,-\right\rangle$ and $\left\vert u'_2,-\right\rangle$ are coupled to the trion state $\left\vert X^-,-\right\rangle=e^{\dag}_{-,\uparrow}e^{\dag}_{-,\downarrow}h^{\dag}_{-,\Uparrow}\left\vert G\right\rangle$ by a $\sigma^-$ polarized light (c.f. Fig. \ref{fig5} (b) and (c)). By virtual excitation of these trion states, single qubit operations including the spin initialization and spin rotations can be controlled via optical Raman process. \cite{atature2006quantum,liu2010quantum} The effective Rabi frequencies for the qubits are the same as the ones in Eq. (\ref{n}) while with $\theta$ being replaced by $\theta'$ here.

$B_x$ plays the same role as intervalley coupling $h$ in the former case discussed in sec. \ref{withvalleyhybridization} and its competition with spin-valley coupling $\lambda$ determines the operation speed.
For MoS$_2$ with $\lambda$ being a few meV, $B_x$ can be the same order of magnitude in a magnetic field of a few Tesla. For the other three group-VIB TMDs, $\lambda \sim 20-40$ meV, which is much larger than $B_x$ in most conditions. The coupling strength of the optical transition from $|u'_1,\tau \rangle $ to $|X^-, \tau \rangle $ in the $\Lambda$-level scheme is a weak one, proportional to $\frac{B_x}{\lambda}$. This will limit the operation speed for the optical control of the states. Note that because of the difference in the effective field $\vec{d'}$, the effective Rabi frequency and hence the operation speed differ  by a factor $\sqrt{\frac{B_x^2+(\lambda/2+B_z)^2}{B_x^2+(\lambda/2-B_z)^2}}$ for the two valleys.

\section{Interplay of lattice nuclear spins with confined electron and hole \label{hyperfine}}

Electrons and holes localized in semiconductors can be coupled to the environment consists of phonons and lattice nuclear spins. At a temperature low for the electrons but high for the nuclear spins (i.e.~10 mK - K), the effects of phonon can be well suppressed, leaving the lattice nuclear spins as the ultimate environmental degrees of freedom.\cite{liu2010quantum} In MX$_2$ nanostructures, the stable isotopes of the relevant elements with nonzero nuclear spin include: ($^{95}$Mo, 5/2, 15.92$\%$), ($^{97}$Mo, 5/2, 9.55$\%$), ($^{183}$W, 1/2, 14.31$\%$), ($^{33}$S, 3/2, 0.76$\%$) and ($^{77}$Se, 1/2, 7.63$\%$), where the second number in the bracket gives the nuclear spin quantum number and the third gives the natural abundance. \cite{web}

We derive here the forms of hyperfine interaction between the localized electron and hole with these lattice nuclear spins in the envelope function approximation. This is applicable for the localized electron wavefunction formed largely from the band edge Bloch functions at the $\pm$K points. These band edge Bloch functions are mainly contributed from the metal d-orbitals and a small but finite component of the chalcogen p-orbitals. The hyperfine coupling strength with the metal nuclear spins are therefore stronger. The chalcogen nuclear spins are expected to play less important roles, for both the weakness of the hyperfine interaction strength and the smaller natural abundance of the stable isotopes with finite nuclear spins. \cite{liu2015electronic,web}

We find that, similarly to both the electron and hole hyperfine interactions in III-V semiconductors, \cite{fischer2008spin,eble2009hole}
the hyperfine interaction here is of the short-range nature: it needs to be counted only for nuclear spins in direct contact with the electron or hole, with a coupling strength proportional to the electron/hole density at the nuclear site. As the electron now has the valley pseudospin in addition to the spin, the hyperfine interaction has intravalley terms as well as the intervalley terms. The latter arises from the short-range nature of the hyperfine interaction, which makes possible the coupling between the single electron states from different valleys.

\subsection{Intravalley and intervalley hyperfine interaction} \label{Hyperfineint}

The hyperfine interaction in the quantum dot is formulated by projecting the complete electron nuclear hyperfine interaction into the basis of the localized electron and hole wavefunctions in the envelope function approximation, which are given by,
\begin{equation}
\Psi_{\tau,s}^{c(v)}(\vec{r})=F^{c(v)}(\vec{r})\Phi_{\tau }^{c(v)}(\vec{r})\chi_{s}
\end{equation}
where $\Phi_{\tau}^{c(v)}(\vec{r})=e^{i \tau \vec{K}\cdot\vec{r}}u_{\tau}^{c(v)}(\vec{r})$ is the Bloch wave function at $\tau K$ point in the conduction (c) and valence (v) bands with $u_{\tau}^{c(v)}(\vec{r})$ being its periodic part,
$F^{c(v)}(\vec{r})$ is the localized envelop function and $\chi_{s}$ is the spin part of the wavefunction.
For electrons, we consider the projected form of the hyperfine interaction between the basis states $\left\{ \Psi_{+,\uparrow}^{c}(\vec{r}),\Psi_{+,\downarrow}^{c}(\vec{r}),\Psi_{-,\uparrow}^{c}(\vec{r}),\Psi_{-,\downarrow}^{c}(\vec{r})\right\} $.
For holes, with the giant spin splitting at the valence
band top, we only need to consider the two-fold spin-valley locked basis: $\left\{\Psi_{+,\uparrow}^{v}(\vec{r}), \Psi_{-,\downarrow}^{v}(\vec{r})\right\}$.

We have used two approaches to obtain the band edge Bloch functions. In the first approach, we extract the orbital compositions of the band edge Bloch states from first principle calculations, and then write the Bloch functions by using the Roothaan-Hartree-Fock atomic orbitals.\cite{Roothaan1951,Roothaan1960} In the second approach, we use the numerically calculated Bloch functions from Abinit.\cite{gonze2002first,gonze2005brief,gonze2009abinit} The two approaches give consistent results on the form and magnitude of hyperfine interactions. Details are given in Appendix B, and the forms are summarized below.
\begin{table}[!hbp]
\begin{tabular}{|c|c|c|c|c|}
\hline
\hline
 & MoS$_2$ & MoSe$_2$ & WS$_2$ & WSe$_2$ \\
\hline
$\varrho$ & 0.23 & 0.23 & 0.37 & 0.40 \\
\hline
$A_M^c$ & -0.50 & -0.51 & 0.76 & 0.79 \\
\hline
$A_M^v$ & -1.52 & -1.53 & 1.78 & 1.82 \\
\hline
$A_X^c$ & 0.05 & 0.46 & 0.08 & 0.33 \\
\hline
$A_X^v$ & -0.16 & -1.33 & -0.37 & -1.63 \\
\hline
\end{tabular}
\caption{Hyperfine constants evaluated based on Bloch functions constructed using Roothaan-Hartree-Fock atomic orbitals (see text). $A_{M(X)}^{c(v)}$ is in unit of $\mu eV$ and $\varrho$ is dimensionless.} \label{HFIpara}
\end{table}

(i) {\it Electron hyperfine interaction with M atom}:
\begin{eqnarray}
H_{M}^{c}&=&A_{M}^c\sum_{k}\Omega|F^c(\vec{R}_{k})|^{2} \left[I^k_{z}S_{z}+\varrho (I^k_{x}S_{x}+I^k_{y}S_{y})\right] \notag \\
&& \times \left(1+e^{-2i\vec{K}\cdot\vec{R}_k}\tau_+ + e^{2i\vec{K}\cdot\vec{R}_k}\tau_-\right), \label{electronwM}
\end{eqnarray}
where $I^{k}$ and $\vec{R}_{k}$ are the spin operator and position vector of the $k$-th nuclei of M atom, $\tau_{\pm}$ are the rasing and lowing operators for valley index and $\Omega$ is the volume of the unit cell. $\varrho$ denotes the ratio between the transverse and the longitudinal interactions.

(ii) {\it Hole hyperfine interaction with M atom}:
\begin{eqnarray}
H^{v}_{M}=A_{M}^v\sum_k \Omega |F^v(\vec{R}_{k})|^{2} I^k_{z}S_{z}. \label{holewM}
\end{eqnarray}

(iii) {\it Electron hyperfine interaction with X atom}:
\begin{eqnarray}
H_{X}^{c}&=&A_{X}^c\sum_{k}\Omega|F^c(\vec{R}_{k}')|^{2} [I'^k_{z}S_{z}+\frac{1}{8} (I'^k_{x}S_{x}+I'^k_{y}S_{y}) \notag \\ &&+\frac{3}{8}(e^{-2i\vec{K}\cdot\vec{R}_k'}\tau_+I'^k_{-}S_-+\text{H.c.})],
\end{eqnarray}
where $I'^{k}$ and $\vec{R}_{k}'$ are the spin operator and position vector of the $k$-th nuclei of X atom and we have used the associated rasing and lowing operators for nuclei and electron spins.

(iv) {\it Hole hyperfine interaction with X atom}:
\begin{eqnarray}
H_{X}^{v}&=&A_{X}^v\sum_{k}\Omega|F^v(\vec{R}_{k}')|^{2} [I'^k_{z}S_{z}-\frac{1}{4}(e^{-2i\vec{K}\cdot\vec{R}_k'}\tau_+I'^k_{+}S_+ \notag \\
&&+\text{H.c.})]. \label{holewX}
\end{eqnarray}
All of the hyperfine constants $A_{M(X)}^{c(v)}$ in different MX$_2$ are listed in Table \ref{HFIpara}.

From the above results, one can find that the hyperfine interaction related to the M nuclei is much stronger then the one to the X nuclei. More importantly, the hyperfine interaction may contain both intravalley and intervalley terms.

\subsection{Optical control of electron-nuclear spin entanglement}\label{entanglement}

The intervalley part in the hyperfine interaction suggests a possibility for optical control of the electron-nuclear spin entanglement. The hyperfine interaction between the confined electron and M nucleus is
\begin{eqnarray}
H_{HF}&=&A_{M}^c\Omega|F^c(0)|^{2} \left[I_{z}S_{z}+\frac{\varrho}{2} (I_{+}S_{-}+I_{-}S_{+})\right] \notag \\
&& \times \left(1+\tau_+ + \tau_-\right). \label{electronwM}
\end{eqnarray}
where we have assumed that the M nuclei is located at position $\vec{R}_0=0$, where the hyperfine interaction is strong. The term $\propto$ $I_zS_z$ shift upwards (downwards) the energy levels with $h_l=\frac{1}{4}A_{M}^c\Omega|F^c(0)|^{2}$when electron and nuclear spins point in the same (opposite) direction. The term $\propto$ $[I_+S_-(\tau_++\tau_-)+\text{H.c.}]$ couples the different valley states where electron and nuclear spins point in the opposite direction, which can be rewritten as
\begin{eqnarray}
H_{1}&=&h_{t}\sigma_x, \label{Hyno}
\end{eqnarray}
where $h_{t}=A^c_M \Omega |F(0)|^2\frac{\varrho}{2}$ and $\sigma$ denotes the Pauli matrices defined in the two-dimensional space spanned by \{$\left\vert+,\uparrow\right\rangle_e \left\vert\downarrow\right\rangle_n$, $\left\vert-,\downarrow\right\rangle_e \left\vert\uparrow\right\rangle_n$\} or \{$\left\vert+,\downarrow\right\rangle_e \left\vert\uparrow\right\rangle_n$, $\left\vert-,\uparrow\right\rangle_e \left\vert\downarrow\right\rangle_n$\} with the subscript e and n denoting electron and nuclear states respectively.
The other terms can be neglected, because they couple those states separated by the spin-valley coupling, which is much larger compared to the hyperfine interaction. The magnitude of $h_t\propto \frac{1}{N}$, where $N=\frac{S}{\Omega}$ is the number of unit cell in the quantum dots of area $S$. For a $^{183}$W nuclei in triangular-shape WS$_2$ quantum dots with $N=100$, we estimate $h_t\sim0.0036 \mu$eV.

\begin{figure}[tbp]
\centering
\includegraphics[width=3.3 in,height=2.5 in]{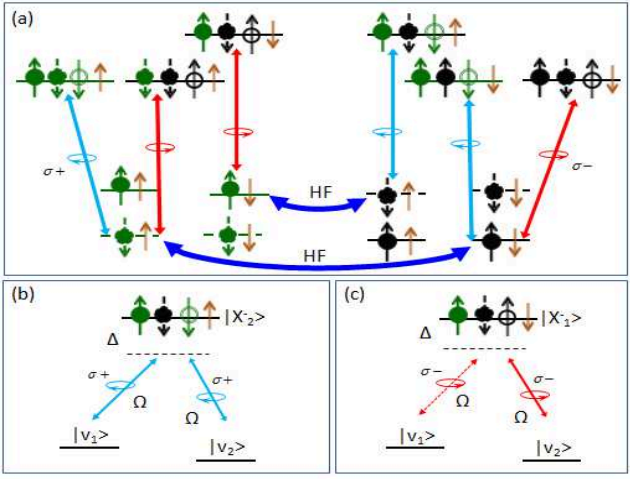}
\caption{(a) Optical and hyperfine couplings between the electron states in presence of a nuclear spin. The brown arrows denote the nuclear spin state, e.g. from a $^{183}$W nuclei. The $\Lambda$-type three-level systems formed by the electron-nuclear entangled states and trion states are connected via an optical pulse with $\sigma^+$ polarization (b) or $\sigma^-$ polarization (c).} \label{fig7}
\end{figure}

The eigenstates of Eq. (\ref{Hyno}), $\left\vert v_{1,2}\right\rangle=\frac{1}{\sqrt{2}}(\left\vert+,\uparrow\right\rangle_e \left\vert\downarrow\right\rangle_n \pm\left\vert-,\downarrow\right\rangle_e \left\vert\uparrow\right\rangle_n)$ or $\left\vert v_{3,4}\right\rangle=\frac{1}{\sqrt{2}}(\left\vert+,\downarrow\right\rangle_e \left\vert\uparrow\right\rangle_n \pm\left\vert-,\uparrow\right\rangle_e \left\vert\downarrow\right\rangle_n)$, which are electron-nuclear entangled states.
These entangled states contain electron spin state from both valleys and can be connected via certain intervalley trion state, as shown in Fig. \ref{fig7} (a). For example, $\left\vert v_{1,2}\right\rangle$ can be coupled with equal strength to the trion state $\left\vert X^-_1\right\rangle=e^{\dag}_{-,\downarrow}e^{\dag}_{+,\uparrow}h^{\dag}_{-,\Uparrow}\left\vert G\right\rangle \left\vert\downarrow\right\rangle_n$ by a $\sigma^-$ polarized light or $\left\vert X^-_2\right\rangle=e^{\dag}_{-,\downarrow}e^{\dag}_{+,\uparrow}h^{\dag}_{+,\Downarrow}\left\vert G\right\rangle \left\vert\uparrow\right\rangle_n$ by a $\sigma^+$ polarized light. Optical Raman processes using these trion states realize an optical quantum pathway to control these electron-nuclear entangled states. However, we note that, because the energy splitting (2$h_{t}$) between $\left\vert v_{1}\right\rangle$ and $\left\vert v_{2}\right\rangle$  is typically less then 1$\mu$eV, the oscillating terms $\propto e^{-i2h_{t}t}$ are slow ones and can not be neglected in this case as we did in Eq. (\ref{Hcouple}). To realize a coherent rotation of the two level system spanned by $\left\vert v_{1}\right\rangle$ and $\left\vert v_{2}\right\rangle$, we use a single optical pulse to couple both states to the trion state, \cite{quinteiro2005entanglement,liu2010quantum} as shown in Fig. \ref{fig7} (b) and (c). Applying an optical pulse with $\sigma^+$ polarization to virtually excite the trion state $\left\vert X^-_2\right\rangle$, the dynamics is governed by the following Hamiltonian,
\begin{equation}
H'=h_{t}\sigma_x-\Delta\left\vert X^-_2 \right\rangle\left\langle X^-_2\right\vert-[\Omega(t)\left\vert -,\downarrow\right\rangle_e \left\vert\uparrow\right\rangle_n\left\langle X^-_2\right\vert+\text{H.c.}], \notag
\end{equation}
where $\Omega(t)$ is the Rabi frequency in the rotating frame and $\Delta$ is the detuning of the laser relative to $\left\vert X^-_2 \right\rangle$. For large detuning,  we can use the adiabatic approximation to eliminate the trion state. The dynamics is described by
\begin{equation}
H'_{eff}=\frac{\left\vert\Omega\right\vert^2}{2\Delta}\left(
    \begin{array}{cc}
      1 & e^{ih_{t}t} \\
      e^{-ih_{t}t} & 1 \\
    \end{array}
  \right)=\epsilon_0 I+\vec{n}(t)\cdot\vec{\sigma}, \label{Heff2}
\end{equation}
describing the qubit state precessing under a time-dependent magnetic field $\vec{n}(t)$ with the strength $\Omega^2/(2\Delta)$ rotating in the x-y plane with the angular frequency $h_{t}$. Because $h_{t}$ is typically less then 1 $\mu$eV and $\Omega^2/(2\Delta)$ is several hundreds of $\mu$eV, the optical pulse in the picosecond scale can be regarded as an instantaneous one. To complete an arbitrary rotation, two subsequent rotations along x- and y-directions, which constitute two SU(2) generators, are needed. Explicitly, at $t=\frac{2n\pi}{h_{t}}$ (n is an integer), $\vec{n}(t)$ is in the x-direction. Whereas at $t=\frac{(2n+1/2)\pi}{h_{t}}$, $\vec{n}(t)$ is in the y-direction. For a $^{183}$W nuclei in triangular-shape WS$_2$ quantum dots with $N=100$, the shortest time interval for these two subsequent operations is 287 ns. Because $h_{t}\propto \frac{1}{N}$, this operation time can be shortened by using smaller quantum dots.

We note that the possibility to control the coherent rotation in the subspace spanned by $\left\vert+,\uparrow\right\rangle_e\left\vert\downarrow\right\rangle_n$ and $\left\vert-,\downarrow\right\rangle_e \left\vert\uparrow\right\rangle_n$, combined with the RF control that flips the nuclear state, can potentially realize the intervalley rotation in the electron subspace spanned by $\left\vert+,\uparrow\right\rangle_e$ and $\left\vert-,\downarrow\right\rangle_e$. We also note that, in our scheme, a single nuclei with state $\left\vert s\right\rangle_n$ is used to couple electron states of different valleys. The interaction strength $h_t\propto\frac{1}{N}$. Alternatively, if a connection of nuclei with a fully polarized initial state $\left\vert\downarrow,\downarrow,\downarrow\cdot\cdot\cdot\right\rangle_n$ are used, the interaction strength $\propto\frac{\sqrt{\nu}}{\sqrt{N}}$, so that the operation speed can be increased by $\sqrt{\nu N}$ time, where $\nu$ is the abundance of M nuclei. In this case, the electron-nuclear entangled states are $\left\vert v'_{1,2}\right\rangle=\frac{1}{\sqrt{2}}(\left\vert+,\uparrow\right\rangle_e \left\vert\downarrow,\downarrow,\downarrow\cdot\cdot\cdot\right\rangle_n \pm\left\vert-,\downarrow\right\rangle_e \sum_k c_k\left\vert\downarrow,\downarrow,\uparrow_k\cdot\cdot\cdot\right\rangle_n)$.

\subsection{Nuclear spin induced decoherence}

The interaction with lattice nuclear spins causes the decoherence of localized electron spin.\cite{merkulov2002electron,khaetskii2002electron,coish2004hyperfine,saykin2002relaxation,witzel2005quantum,yao2006theory} Because the hyperfine
interaction with the M nuclei is much stronger than the one with the X nuclei, we consider the decoherence effect arising from interaction with the former one. From Eq. (\ref{electronwM}), we know that there are four decoherence channels in the basis of $\{\left\vert +,\uparrow\right\rangle,\left\vert -,\downarrow\right\rangle,\left\vert +,\downarrow\right\rangle,\left\vert -,\uparrow\right\rangle\}$. The first one arises from the term $\propto$ $I^k_zS_z$, which causes dephasing between electron states with different spin. The second one arises from the term $\propto$ $[I^k_+S_-(\tau_++\tau_-)+\text{H.c.}]$, which causes relaxation between electron states with different spin and valley. The third one arises from the term $\propto$ $[I^k_zS_z(\tau_++\tau_-)+\text{H.c.}]$, which causes relaxation between electron states with different valley while the same spin. The last one arises from the term $\propto$ $(I^k_+S_-+\text{H.c.})$, which causes relaxation between electron states with different spin in each valley. The last two relaxation channels are much weaker compared with the former two, because the hyperfine interaction is small compared to the spin-valley coupling so that the energy cost associated with the transitions (a few to a few tens meV) is much larger than the hyperfine induced transition matrix element. So relaxation between the initial and final states are suppressed by the large energy cost.  Therefore, in the following, we make an estimation of the decoherence time arising from the first two channels.

For the dephasing between electron states
with different spin induced by the term $\propto$ $I^k_zS_z$, the effective nuclear field experienced by the localized electron in each single valley is $h_{eff}=A_{M}^c\sum_{k}\Omega|F_Q^c(\vec{R}_{k})|^{2}I^k_{z}$. The statistical fluctuation in the nuclear spin configurations therefore corresponds to an uncertainty in the energy difference between the electron states with opposite spins, and hence results in pure (inhomogeneous) dephasing. We assume that there is no correlation between different nuclear spins and that the nuclear spins are distributed uniformly within the quantum dot, the variance of the field is
\begin{equation}
\left\langle h_{eff}^2 \right\rangle=\left(A_{M}^c\right)^2\sum_{k}\nu \Omega^2|F_Q^c(\vec{R}_{k})|^{4}\left\langle I^{k2}_{z} \right\rangle, \label{dec2}
\end{equation}
where $\left\langle\cdots\right\rangle$ denotes the average over nuclear spin states.
The coherence time for electron state is $T^*_2\sim\frac{2\pi}{\sqrt{\langle h_{eff}^2 \rangle}}$.

The intervalley electron-nuclear flip-flop term $\propto$ $[I^k_+S_-(\tau_++\tau_-)+\text{H.c.}]$ causes the population relaxation between the degenerate electron states $|+,\uparrow>$ and $|-,\downarrow>$ and between $|+,\downarrow>$ and $|-,\uparrow>$. The relaxation time for this process is $T_1\sim\frac{2\pi}{\sqrt{\left\langle h'^2_{eff} \right\rangle}}$,\cite{merkulov2002electron} where $\left\langle h'^2_{eff} \right\rangle$ is the variance of the in-plane nuclear field,
\begin{equation}
\left\langle h'^2_{eff} \right\rangle=\varrho^2\left(A_{M}^c\right)^2\sum_{k}\nu \Omega^2|F_Q^c(\vec{R}_{k})|^{4}\left\langle (I^{k2}_{x}+I^{k2}_{y}) \right\rangle. \label{dec1}
\end{equation}

\begin{table}[!hbp]
\begin{tabular}{|*{8}{c|}}
\hline
&\multicolumn{1}{c|}{$N$} & \multicolumn{2}{c|}{$100$} & \multicolumn{2}{c|}{$500$} & \multicolumn{2}{c|}{$1000$} \\
\hline
\multirow{3}{*}{{\shortstack{MoS$_2$}}}&$\nu$ & $25.5\%$ & $100\%$ & $25.5\%$ & $100\%$ & $25.5\%$ & $100\%$ \\
\cline{2-8}
&$T_1(ns)$ & 298 & 150 & 668 & 337 & 944 & 477 \\
\cline{2-8}
&$T_2^*(ns)$ & 97 & 49 & 217 & 110 & 307 & 155 \\
\hline
\multirow{3}{*}{{\shortstack{WS$_2$}}}&$\nu$ & $14.3\%$ & $100\%$ & $14.3\%$ & $100\%$ & $14.3\%$ & $100\%$ \\
\cline{2-8}
&$T_1(ns)$ & 557 & 210 & 1246 & 471 & 1762 & 666 \\
\cline{2-8}
&$T_2^*(ns)$ & 292 & 110 & 652 & 246 & 922 & 349 \\
\hline
\end{tabular}
\caption{Decoherence time in MoS$_2$ and WS$_2$ quantum dots with different numbers of nuclear spin. $\nu=25.5\%$ $(14.3\%)$ corresponds to the natural abundance in MoS$_2$ (WS$_2$), and $\nu=100\%$ corresponds to the case that each metal atom within the quantum dot has a nuclear spin.} \label{detime}
\end{table}
For an infinite-temperature state, we have $\left\langle I^{k2}_{x,y,z}\right\rangle=I^k(I^k+1)/3$. Since $\sum_{k}\Omega^2|F_Q^c(\vec{R}_{k})|^{4}\sim\frac{1}{N}$, the decoherence time increases with the increase of $\sqrt{N}$. In Table \ref{detime}, we list $T_2^*$ and $T_1$ for quantum dots with different size (represented by $N$) and the abundance of the nuclear spins. The decoherence time is several hundreds of ns, which is serval orders larger then the operation time in the optical quantum control of spin-valley qubit.

\begin{figure}[tbp]
\includegraphics[width=3.3 in,height=2.5 in]{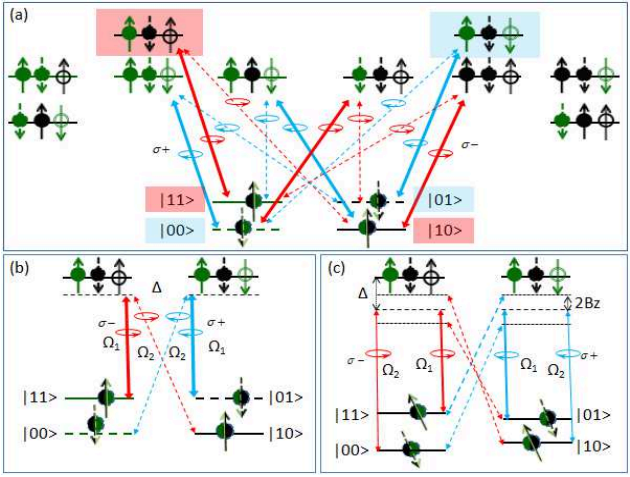}
\caption{Optically controlled two-qubit operations between the spin qubit and valley qubit carried by a single electron in a quantum dot. (a) Definition of the two-qubit states. The trion states with higher energy (highlighted in red and blue background) are used as intermediate states for optical control in two-qubit operations. (b) Energy level scheme for realizing controlled phase gate and controlled NOT gate in the presence of valley hybridization. (c) Energy level scheme for realizing SWAP gate in the presence of valley hybridization and applied magnetic fields. The unwanted transitions (dashed doubled-arrowed lines) are detuned from the two-photon resonant condition by 2$B_z$ and suppressed.}\label{add}
\end{figure}

\section{Optically controlled two-qubit operations in a single quantum dot}\label{addsection}
With the extra valley degree of freedom in TMDs, a single electron in the ground state of the QD confinement has four spin-valley configurations that one can exploit to encode two qubits. Here we discuss the possibility of utilizing both qubits in a single dot and realizing two-qubit logic controls.

We consider first the logic operations in the presence of valley hybridization, but in the absence of in-plane magnetic field. We define the computational basis as $\left\vert mn\right\rangle=\left\vert m\right\rangle_s\otimes\left\vert n\right\rangle_v$ where $m,n=\{1,0\}$, and the subscripts $s$ and $v$ denote spin and valley degrees of freedom. Explicitly, $\left\vert11\right\rangle=\left\vert u_1,\uparrow\right\rangle$, $\left\vert10\right\rangle=\left\vert u_2,\uparrow\right\rangle$, $\left\vert01\right\rangle=\left\vert u_1,\downarrow\right\rangle$, and $\left\vert00\right\rangle=\left\vert u_2,\downarrow\right\rangle$ (see Fig. \ref{add} (a)). Under this definition, the two qubits do not have interaction at rest. The optical control for two-qubit operations uses the two trion states $e^{\dag}_{-,\downarrow}e^{\dag}_{+,\uparrow}h^{\dag}_{-,\Uparrow}\left\vert G\right\rangle$ and $e^{\dag}_{-,\downarrow}e^{\dag}_{+,\uparrow}h^{\dag}_{+,\Downarrow}\left\vert G\right\rangle$ which have energies higher than the other trion states by the conduction band spin-orbit splitting $\lambda$. As highlighted in Fig. \ref{add} (a), these two trion states couple to the four states of the two qubits by light of different circular polarizations. These two trion states then can be used as the intermediate states in our control scheme, where the lower energy trion states can be neglected with $\lambda$ in the range of a few meV to a few tens of meV \cite{liu2013three,liu2015electronic} (c.f. Fig. \ref{add} (b) and (c)). Via virtually exciting these trion states with different circularly polarized lights, one can obtain the controlled two-qubit gates. For example, applying a single pulse of $\sigma^+$ light ($\Omega_1=0$), only the state $\left\vert00\right\rangle$ will pick up a phaseshift due to the AC-stark shift (Fig. \ref{add} (b)), so we have a controlled phase-shift gate
\begin{equation}
U_{P_{\phi}}=\left(
      \begin{array}{cccc}
        1 & 0 & 0 & 0 \\
        0 & 1 & 0 & 0 \\
        0 & 0 & 1 & 0 \\
        0 & 0 & 0 & e^{i\phi} \\
      \end{array}
    \right).
\end{equation}
One can also use two $\sigma^+$ polarized pulses to selectively couple the $\left\vert00\right\rangle$ and $\left\vert01\right\rangle$ states via a Raman-type process (c.f. Fig. \ref{add} (b)). By controlling the amplitudes and phases of the two pulses so that the pseudomagnetic field $\vec{n}$ defined in Eq. (\ref{n}) is in the $x$ direction, a controlled NOT gate can be realized
\begin{equation}
U_{N}=\left(
      \begin{array}{cccc}
        1 & 0 & 0 & 0 \\
        0 & 1 & 0 & 0 \\
        0 & 0 & 0 & 1 \\
        0 & 0 & 1 & 0 \\
      \end{array}
    \right).
\end{equation}
For MoS$_2$ quantum dot with an intervalley coupling strength of 1 meV (see Appendix A), if we set the detuning $\Delta=0.3$ meV and $E_1 D_0=0.05$ meV, this two-qubit gate can be realized in $\sim$ 5 ns.

To realize a SWAP gate, we consider nanostructures with valley hybridization and in applied external magnetic field. In this scenario, the electron eigenstates are both spin and valley hybridized, which are used as the basis states for the qubits,
\begin{eqnarray}
\left\vert11\right\rangle&=&(\cos\theta\left\vert +\right\rangle+\sin\theta\left\vert -\right\rangle)(\cos\theta'\left\vert \uparrow\right\rangle+\sin\theta'\left\vert \downarrow\right\rangle), \notag \\
\left\vert00\right\rangle&=&(\cos\theta\left\vert +\right\rangle+\sin\theta\left\vert -\right\rangle)(\sin\theta'\left\vert \uparrow\right\rangle-\cos\theta'\left\vert \downarrow\right\rangle), \notag \\
\left\vert01\right\rangle&=&(\sin\theta\left\vert +\right\rangle-\cos\theta\left\vert -\right\rangle)(\sin\theta'\left\vert \uparrow\right\rangle-\cos\theta'\left\vert \downarrow\right\rangle), \notag \\
\left\vert10\right\rangle&=&(\sin\theta\left\vert +\right\rangle-\cos\theta\left\vert -\right\rangle)(\cos\theta'\left\vert \uparrow\right\rangle+\sin\theta'\left\vert \downarrow\right\rangle). \notag
\end{eqnarray}
In this definition, the two qubits do not have interaction at rest. Each of these states is now optically coupled to the trion states $e^{\dag}_{-,\downarrow}e^{\dag}_{+,\uparrow}h^{\dag}_{-,\Uparrow}\left\vert G\right\rangle$ and $e^{\dag}_{-,\downarrow}e^{\dag}_{+,\uparrow}h^{\dag}_{+,\Downarrow}\left\vert G\right\rangle$ (c.f. Fig. \ref{add} (c)). For example, applying $\sigma^+$ polarized lights, these states can couple to the trion state $e^{\dag}_{-,\downarrow}e^{\dag}_{+,\uparrow}h^{\dag}_{+,\Downarrow}\left\vert G\right\rangle$ with strengths $\propto$ $\sin\theta\sin\theta'$, $\sin\theta\cos\theta'$, $\cos\theta\cos\theta'$ and $\cos\theta\sin\theta'$ respectively. In order to selectively control these states, we apply a magnetic field in $z$ direction to make the unwanted optical transitions detuned from the two-photon resonant condition by the Zeeman splitting $2B_z$ and suppressed, as shown in Fig. \ref{add} (c). Via virtually exciting the trion state $e^{\dag}_{-,\downarrow}e^{\dag}_{+,\uparrow}h^{\dag}_{+,\Downarrow}\left\vert G\right\rangle$ with $\sigma^+$ light, one can realize coherent rotations selectively between $\left\vert01\right\rangle$ and $\left\vert10\right\rangle$ to realize a SWAP gate,
\begin{equation}
U_{W}=\left(
      \begin{array}{cccc}
        1 & 0 & 0 & 0 \\
        0 & 0 & 1 & 0 \\
        0 & 1 & 0 & 0 \\
        0 & 0 & 0 & 1 \\
      \end{array}
    \right).
\end{equation}
For MoS$_2$ quantum dot with an intervalley coupling strength of 1 meV (see Appendix A) and applied magnetic fields $B_x$=$B_z$=1 meV, if we set the detuning $\Delta=0.3$ meV and $E_1 D_0=0.05$ meV, the SWAP gate can be realized in $\sim$ 2 ns.

\section{Discussion and Conclusions \label{Conclusions}}

In conclusion, we have studied the optical controllability of the spin-valley qubit carried by single electrons localized in nanostructures of monolayer TMDs, including small quantum dots and charged impurities. Various control scenarios with and without valley hybridization caused by the quantum confinement are considered. For nanostructures with finite intervalley coupling, the low-energy states are valley-hybridized with definite spin index. Because of valley hybridization, the electron states with the same spin index can be coupled to a common trion state by lights of the same polarization, which makes possible the arbitrary coherent rotation via Raman processes. And states with different spin index can be selectively accessed by light of different circular polarization. Without intervalley coupling, which is the case when the confinements have C$_3$ or higher rotational symmetry about a chalcogen atom site or the hollow center of the hexagon lattice, we use a magnetic field with an in-plane component to hybridize the spin states in each valley. The low-energy states in this case are spin-hybridized with a definite valley index and can also be selectively accessed by light of different circular polarization. For a single electron confined in the nanostructure, its four spin-valley configurations can encode two qubits, where two-qubit logic operations such as the controlled NOT gate, controlled phase gate, and SWAP gate can be realized in the presence of the valley hybridization.

We also studied the effect of interaction with lattices nuclear spins on the localized electrons and holes in the nanostructures. The hyperfine interaction has intervalley terms besides intravalley ones, because of its short-range nature. Based on this, we studied the possibility to optically control the electron-nuclear spin entanglement. Some decoherence channels induced by the statistical fluctuations of the nuclear spin configurations are discussed.

Controlled interplay between electrons localized in adjacent nanostructures may be realized using schemes developed for coupling III-V quantum dots, e.g. by applying an electrical gate to tune the tunneling amplitude between two dots \cite{loss1998quantum} or virtually exciting the delocalized exciton to interact with the electrons in both dots \cite{piermarocchi2002optical}. The generalization and quantitative analysis of these schemes in TMDs nanostructures will be interesting topics for future studies.

We also note that the breaking of mirror symmetry about the metal atom plane can give rise to Rashba-type spin-orbit coupling which, together with phonon scattering, can be an important cause for the relaxation of spin-valley qubit. \cite{kormanyos2014spin} This can be the case for quantum dots defined by patterned electrodes \cite{kormanyos2014spin} or charged impurity at a chalcogen atom site. The detailed investigation of the mechanisms and timescales for the relaxation and decoherence of spin-valley qubit in systems with or without mirror symmetry is also an interesting topic for future study.

\section{ACKNOWLEDGMENTS}

The work was supported by the Croucher Foundation (Croucher Innovation Award), the Research Grant Council of HKSAR (HKU705513P, HKU9/CRF/13G), and the HKU OYRA and ROP. GBL acknowledges the support by NSF of China (No. 11304014).

\appendix

\section{Intervalley coupling strength in the confinement by small quantum dots and charged impurity \label{intervalleycoupling}}

We use the real-space tight-binding (RSTB) method to calculate the intervalley coupling strength in different types of quantum dots as well as the charged impurity systems. The validity of this RSTB method in calculating the intervalley coupling has been tested by comparing it with an entirely different approach, i.e. the envelope function method as discussed in Ref. \cite{liu2014intervalley}. We calculate the strength of intervalley coupling in the quantum dots and impurity systems with supercells and using periodic boundary conditions.

\begin{figure}[tbp]
\includegraphics[width=1\columnwidth]{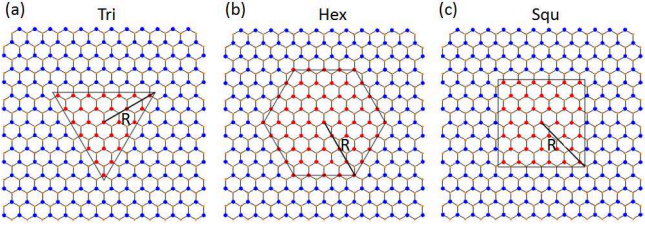}
\caption{Schematics of quantum dot confinement potential with triangular (a), hexagonal (b), and square (c) shapes. The red spots denote quantum dot regime and the potentials are all centered at M atoms. $R$ is the potential radius.}
\label{RofQD}
\end{figure}

\begin{figure}[tbp]
\includegraphics[width=3.5 in,height=2.0 in]{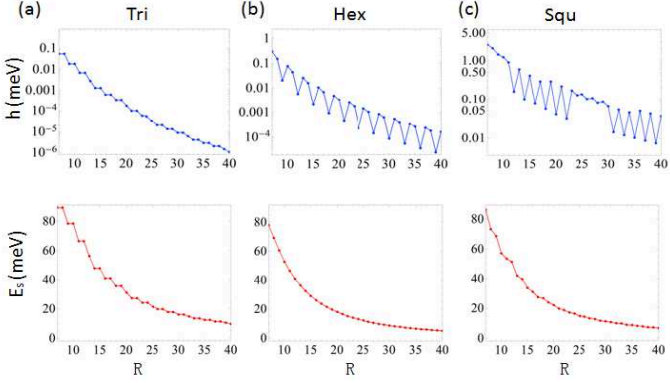}
\caption{Intervalley coupling strength $h$ and the energy separation $E_s$ between the first excited state and ground state as functions of $R$ in MoS$_2$ quantum dots of triangular (a), hexagonal (b) and square (c) shapes. We set the confinement potential to be $0.2$ eV and take the length unit as the lattice constant $a=3.193\mathring{A}$.}
\label{app2}
\end{figure}

To calculate the strength of intervalley coupling in small quantum dots, we take monolayer MoS$_2$ system with lateral confinement as an example. We consider three different types of confinement potential, i.e. the triangular-shape, hexagonal-shape and square-shape potentials. The center of the confinement potential is set at the Mo site to maximize the intervalley coupling (c.f. Fig. \ref{RofQD}). The results are plotted in Fig. \ref{app2}. Under an external confinement of $0.2 eV$, the intervalley coupling is on the scale of $meV$ in small quantum dots (several lattice length scale). With the increase of the potential radius $R$, the intervalley coupling decreases. In triangular and hexagonal quantum dots, the intervalley coupling decreases very fast to $\mu eV$ order and even lower when $R$ increases. However in square quantum dots with the same $R$ the coupling strength is still very large. We also calculate the energy separation between the ground state and the first excited state in the quantum dots. We find that the energy separation also decreases with the increase of $R$. For all three types of quantum dots with $R=7\sim30a$, with $a$ being the lattice constant, the energy separation is about $10\sim90 meV$, which convinces us that the excite-state levels are well separated from the ground-state level.

\begin{table}[!hbp]
\begin{tabular}{ccccc}
\hline
\hline
$U_{on}(eV)$ & -0.5 & -1.0 & -1.0 & -1.0 \\
$\epsilon_r$ & 16 & 16 & 7 & 27 \\
$h(meV)$ & 2 & 14 & 4 & 10 \\
$E_s(meV)$ & 49 & 49 & 169 & 19 \\
\hline
\hline\label{tableapp}
\end{tabular}
\caption{Intervalley coupling strength $h$ and the energy separation between the first excited state and ground state $E_s$ in WSe$_2$ impurity system for different parameters including the on-site confinement $U_{on}$ and relative dielectric constant $\epsilon_r$.} \label{ImpurityVC}
\end{table}
To calculate the strength of intervalley coupling in charged impurity systems, we consider the example of monolayer WSe$_2$ with one W atom replaced by a Re one, using the 2D hydrogenic confinement potential $-\frac{e^2}{4\pi \epsilon_r \epsilon_0 r}$, where $r$ is the distance from the impurity center, and $\epsilon_r$ is the relative dielectric constant. The known corrections to the Coulomb potential, and the possible break down of the effective mass approximation for the donor system will make the quantitative numbers here inaccurate. However, we want to give some estimation of the order of magnitude of the intervalley coupling strength for the strongly localized electron. We use the estimated exciton binding energy to describe the on-site confinement $U_{on}$ at the impurity cite, i.e. about $-1.0\sim-0.5 eV$. The calculated intervalley coupling strength $h$ and the energy separation between first excited state and ground state $E_s$ are listed in table III. From the results, one can see that the first excited state are well separated from the ground state and can be safely neglected.

\section{Derivation of hyperfine interaction \label{hyperfinedetail}}
The complete form of the hyperfine interaction between the electron and nuclear
spins is
\begin{eqnarray}
H_{hf}&=&-\frac{\mu_{0}\gamma_{e}}{4\pi}\sum_{k}\gamma_{k}\bm{I}_{k}\cdot[\frac{\bm{L}}{r_{k}^{3}}-\frac{r_{k}^{2}\bm{S}-3\bm{r}_{k}(\bm{S}\cdot\bm{r}_{k})}{r_{k}^{5}} \notag \\
&&+\frac{8\pi}{3}\bm{S}\delta(\bm{r}_{k})], \label{HhfA}
\end{eqnarray}
where $\mu_{0}$ is the vacuum permeability, $\gamma_{e}$ and $\gamma_{k}$
are the gyromagnetic ratios of electron and nuclei, $\bm{r}_{k}=\bm{r}-\bm{R}_{k}$
is the electron coordinate measured from the $k$-th nucleus and $\bm{S}$
and $\bm{L}$ are the spin and angular momentums of the electron
respectively.

As one may directly infer from Eq. (\ref{HhfA}), in order to get reduced form of the hyperfine interaction for the localized electron and hole in the envelope function approximation, it is important to study several integrals concerning
$\bm{r}_{k}$ for the band edge Bloch functions. Because there are two species of nuclei in the system, we use $\vec{R}_k$ to denote the position of Mo nuclei and $\vec{R}_k'$ for S nuclei (accordingly we have $\vec{r}_k=\vec{r}-\vec{R}_k$ and $\vec{r}_k'=\vec{r}-\vec{R}_k'$). We have used two approaches to give the band edge Bloch functions. In the first approach, we extract the orbital compositions of the band edge Bloch states from first-principles calculations, and then write the Bloch functions by using the Roothaan-Hartree-Fock atomic orbitals (Appendix B1). In the second approach, we use the numerically calculated Bloch functions from Abinit (Appendix B2). The two approaches give consistent results on the form and magnitude of hyperfine interactions (Eq. (14-17) in main text), which are also consistent with the symmetry analysis presented in Appendix B4. We take MoS$_2$ as an example and list the numerical results from the two approaches below.

\subsection{Evaluation based on Bloch functions constructed using Roothaan-Hartree-Fock atomic orbitals}

\begin{table}[!hbp]
\begin{tabular}{|c|c|c|c|c|c|c|}
\hline
\hline
 & Mo-5s & Mo-4d$_0$ & Mo-4d$_{+2}$ & Mo-4d$_{-2}$ & S-3p$_{+1}$ & S-3p$_{-1}$ \\
\hline
c(+K) & 4.7\% & 87.6\% & 0 & 0 & 7.7\% & 0 \\
\hline
c(-K) & 4.7\% & 87.6\% & 0 & 0 & 0 & 7.7\% \\
\hline
v(+K) & 0 & 0 & 84.3\% & 0 & 0 & 15.7\% \\
\hline
v(-K) & 0 & 0 & 0 & 84.3\% & 15.7\% & 0 \\
\hline
\end{tabular}
\caption{Orbital compositions of the Bloch states in conduction and valence bands for $\pm$K valleys from first principle calculations \cite{liu2015electronic}.} \label{compo}
\end{table}
The Roothaan-Hartree-Fock method gives analytic wave functions for the various orbitals of neutral atoms. Together with the first-principles calculated orbital compositions of both conduction and valence band edge states, as listed in Table \ref{compo}, we are able to give an analytical expression for the band edge Bloch functions for the evaluation of the hyperfine interaction, assuming that the atomic orbitals in the crystal has not changed too significantly from that in the neutral atoms. Similar approach has been used for the evaluation of hyperfine interaction of holes in III-V semiconductors \cite{fischer2008spin}.
In the Roothaan-Hartree-Fock method, Slater-type orbitals \cite{Slater1930} are linearly combined to form the atomic orbitals. The radial part of the Slater-type orbitals is
\begin{equation}
f(r)=N_sr^{n-1}e^{-\zeta r},
\end{equation}
where $N_s$ is a normalization constant and $n$ is the principal quantum number. According to Table \ref{compo}, we focus on three atomic orbitals, Mo-5s, Mo-4d and S-3p. The optimized atomic orbitals for neutral Mo and S atoms are listed in Appendix \ref{orbital}. Based on these results, we write the Bloch functions for the conduction and valence band at $+K$ valley as
\begin{widetext}
\begin{equation}
\begin{array}{rl}
\displaystyle\phi^c_{+}(\vec{r})=&\displaystyle\sum_{\vec{R}_k} e^{i\vec{K}\cdot\vec{R}_k}\left[\alpha_s^c f_{5s}(r_k)Y^0_0(\theta_k,\varphi_k) + \alpha_d^c f_{4d}(r_k)Y^0_2(\theta_k,\varphi_k)\right] + \sum_{\vec{R}_k'}e^{i\vec{K}\cdot\vec{R}_k'}\frac{\alpha_p^c}{\sqrt2}f_{3p}(r_k')Y^{+1}_1(\theta_k',\varphi_k') ,\\[2ex]
\displaystyle\phi^v_{+}(\vec{r})=&\displaystyle\sum_{\vec{R}_k}e^{i\vec{K}\cdot\vec{R}_k}\alpha_d^v f_{4d}(r_k)Y^{+2}_2(\theta_k,\varphi_k) + \sum_{\vec{R}_k'}e^{i\vec{K}\cdot\vec{R}_k'}\frac{\alpha_p^v}{\sqrt2}f_{3p}(r_k')Y^{-1}_1(\theta_k',\varphi_k') , \\[2ex]
\end{array}
\label{Bloch}
\end{equation}
\end{widetext}
where $\alpha^2$ is the orbital composition as listed in Table \ref{compo}, $f_{3p}$, $f_{4d}$, and$f_{5s}$ are the radial parts of the atomic orbitals (see Appendix \ref{orbital}), $Y^0_0$, $Y^0_2$ and $Y^{+1}_1$ are the corresponding spherical harmonics. There is a factor $\frac{1}{\sqrt2}$ before the S-3p orbital part because we have 2 S atoms with mirror symmetry in one unit cell. For the $-K$ valley, we have $\phi^c_{-}(\vec{r})=[\phi^c_{+}(\vec{r})]^*$ and $\phi^v_{-}(\vec{r})=[\phi^v_{+}(\vec{r})]^*$. In this way, we give an estimation of the Bloch wave function $\Psi^{c(v)}_{\tau}(\vec{r})=\phi^{c(v)}_{\tau}(\vec{r})$. Then we can calculate those integrals for the terms involving $r_k$ ($r_k'$) in the Hamiltonian (\ref{HhfA}), and the results are listed in Table \ref{data1} and \ref{data2}. These integrals then lead
to the expressions of the hyperfine interaction in Eqs. (14-17).
In the evaluations of those integrals, we find that only the on-site atomic orbitals have significant contributions to the hyperfine interaction.
Namely,
\begin{equation}
\langle+|\frac{1}{r^3_k}|+\rangle_c=\int_V\! d\vec{r} [\phi^c_+(\vec{r})]^*\frac{1}{r^3_k}\phi^c_+(\vec{r}), \label{onsite}
\end{equation}
where $V$ can be just taken as the unit cell centered at $r_k=0$ (c.f. Fig. \ref{neighborsFigure}). The corrections from nearest neighbor and next nearest neighbor unit cells are found to be negligible.
The same is true for other integrals presented in Table \ref{data1} and Table \ref{data2}. Therefore, although the hyperfine interaction is dominated by the dipolar part for the p and d orbitals, it is still of an ``on-site'' or ``contact'' form. This is similar to the case of the hyperfine interaction for holes in III-V semiconductors as shown in Refs. \cite{fischer2008spin} and \cite{eble2009hole}.

\begin{widetext}

\begin{table}[!hbp]
\begin{tabular}{|c|c|c|c|c|c|c|c|}
\hline
\hline
 & $\langle+|\frac{1}{r^3_k}|+\rangle_c$ & $\langle+|\frac{r^2_{kz}}{r^5_k}|+\rangle_c$ & $\langle+|\frac{r^2_{kx}}{r^5_k}|+\rangle_c$ & $\langle+|\frac{r^2_{ky}}{r^5_k}|+\rangle_c$ & $\langle+|\frac{r_{kx}r_{ky}}{r^5_k}|+\rangle_c$ & $\langle+|\frac{r_{kx}r_{kz}}{r^5_k}|+\rangle_c$ & $\langle+|\frac{r_{ky}r_{kz}}{r^5_k}|+\rangle_c$ \\
\hline
\textit{Abinit} & 24.23 & 11.68 & 6.28 & 6.27 & 0.01 & 0.00 & 0.00 \\
\hline
\textit{RHF} & 20.60 & 10.79 & 4.90 & 4.90 & 0.00 & 0.00 & 0.00 \\
\hline
 & $|\langle+|\frac{1}{r^3_k}|-\rangle_c|$ & $|\langle+|\frac{r^2_{kz}}{r^5_k}|-\rangle_c|$ & $|\langle+|\frac{r^2_{kx}}{r^5_k}|-\rangle_c|$ & $|\langle+|\frac{r^2_{ky}}{r^5_k}|-\rangle_c|$ & $|\langle+|\frac{r_{kx}r_{ky}}{r^5_k}|-\rangle_c|$ & $|\langle+|\frac{r_{kx}r_{kz}}{r^5_k}|-\rangle_c|$ & $|\langle+|\frac{r_{ky}r_{kz}}{r^5_k}|-\rangle_c|$ \\
\hline
\textit{Abinit} & 24.18 & 11.68 & 6.25 & 6.25 & 0.00 & 0.00 & 0.00 \\
\hline
\textit{RHF} & 20.60 & 10.79 & 4.90 & 4.90 & 0.00 & 0.00 & 0.00 \\
\hline
\hline
 & $\langle+|\frac{1}{r^3_k}|+\rangle_v$ & $\langle+|\frac{r^2_{kz}}{r^5_k}|+\rangle_v$ & $\langle+|\frac{r^2_{kx}}{r^5_k}|+\rangle_v$ & $\langle+|\frac{r^2_{ky}}{r^5_k}|+\rangle_v$ & $\langle+|\frac{r_{kx}r_{ky}}{r^5_k}|+\rangle_v$ & $\langle+|\frac{r_{kx}r_{kz}}{r^5_k}|+\rangle_v$ & $\langle+|\frac{r_{ky}r_{kz}}{r^5_k}|+\rangle_v$ \\
\hline
\textit{Abinit} & 19.06 & 2.88 & 8.10 & 8.08 & 0.02 & 0.00 & 0.00 \\
\hline
\textit{RHF} & 19.82 & 2.83 & 8.50 & 8.50 & 0.00 & 0.00 & 0.00 \\
\hline
 & $|\langle+|\frac{1}{r^3_k}|-\rangle_v|$ & $|\langle+|\frac{r^2_{kz}}{r^5_k}|-\rangle_v|$ & $|\langle+|\frac{r^2_{kx}}{r^5_k}|-\rangle_v|$ & $|\langle+|\frac{r^2_{ky}}{r^5_k}|-\rangle_v|$ & $|\langle+|\frac{r_{kx}r_{ky}}{r^5_k}|-\rangle_v|$ & $|\langle+|\frac{r_{kx}r_{kz}}{r^5_k}|-\rangle_v|$ & $|\langle+|\frac{r_{ky}r_{kz}}{r^5_k}|-\rangle_v|$ \\
\hline
\textit{Abinit} & 0.04 & 0.00 & 0.10 & 0.08 & 0.10 & 0.00 & 0.00 \\
\hline
\textit{RHF} & 0.00 & 0.00 & 0.00 & 0.00 & 0.00 & 0.00 & 0.00 \\
\hline
\end{tabular}
\caption{Intravalley and intervalley integrals for the Mo nucleus regime. \textit{Abinit} means the results are from the Abinit wave function, and \textit{RHF} means the Roothaan-Hartree-Fock wave function. The unit used here is $\mathring{A}^{-3}$.} \label{data1}
\end{table}

\begin{table}[!hbp]
\begin{tabular}{|c|c|c|c|c|c|c|c|}
\hline
\hline
 & $\langle+|\frac{1}{r_k'^3}|+\rangle_c$ & $\langle+|\frac{r_{kz}'^2}{r_k'^5}|+\rangle_c$ & $\langle+|\frac{r_{kx}'^2}{r_k'^5}|+\rangle_c$ & $\langle+|\frac{r_{ky}'^2}{r_k'^5}|+\rangle_c$ & $\langle+|\frac{r_{kx}'r_{ky}'}{r_k'^5}|+\rangle_c$ & $\langle+|\frac{r_{kx}'r_{kz}'}{r_k'^5}|+\rangle_c$ & $\langle+|\frac{r_{ky}'r_{kz}'}{r_k'^5}|+\rangle_c$ \\
\hline
\textit{Abinit} & 2.26 & 0.46 & 0.91 & 0.89 & 0.01 & -0.01 & -0.01 \\
\hline
\textit{RHF} & 1.26 & 0.25 & 0.50 & 0.50 & 0.00 & 0.00 & 0.00 \\
\hline
 & $|\langle+|\frac{1}{r_k'^3}|-\rangle_c|$ & $|\langle+|\frac{r_{kz}'^2}{r_k'^5}|-\rangle_c|$ & $|\langle+|\frac{r_{kx}'^2}{r_k'^5}|-\rangle_c|$ & $|\langle+|\frac{r_{ky}'^2}{r_k'^5}|-\rangle_c|$ & $|\langle+|\frac{r_{kx}'r_{ky}'}{r_k'^5}|-\rangle_c|$ & $|\langle+|\frac{r_{kx}'r_{kz}'}{r_k'^5}|-\rangle_c|$ & $|\langle+|\frac{r_{ky}'r_{kz}'}{r_k'^5}|-\rangle_c|$ \\
\hline
\textit{Abinit} & 0.00 & 0.01 & 0.45 & 0.44 & 0.45 & 0.01 & 0.01 \\
\hline
\textit{RHF} & 0.00 & 0.00 & 0.25 & 0.25 & 0.25 & 0.00 & 0.00 \\
\hline
\hline
 & $\langle+|\frac{1}{r_k'^3}|+\rangle_v$ & $\langle+|\frac{r_{kz}'^2}{r_k'^5}|+\rangle_v$ & $\langle+|\frac{r_{kx}'^2}{r_k'^5}|+\rangle_v$ & $\langle+|\frac{r_{ky}'^2}{r_k'^5}|+\rangle_v$ & $\langle+|\frac{r_{kx}'r_{ky}'}{r_k'^5}|+\rangle_v$ & $\langle+|\frac{r_{kx}'r_{kz}'}{r_k'^5}|+\rangle_v$ & $\langle+|\frac{r_{ky}'r_{kz}'}{r_k'^5}|+\rangle_v$ \\
\hline
\textit{Abinit} & 4.14 & 0.79 & 1.68 & 1.67 & 0.00 & 0.00 & 0.00 \\
\hline
\textit{RHF} & 2.57 & 0.51 & 1.03 & 1.03 & 0.00 & 0.00 & 0.00 \\
\hline
 & $|\langle+|\frac{1}{r_k'^3}|-\rangle_v|$ & $|\langle+|\frac{r_{kz}'^2}{r_k'^5}|-\rangle_v|$ & $|\langle+|\frac{r_{kx}'^2}{r_k'^5}|-\rangle_v|$ & $|\langle+|\frac{r_{ky}'^2}{r_k'^5}|-\rangle_v|$ & $|\langle+|\frac{r_{kx}'r_{ky}'}{r_k'^5}|-\rangle_v|$ & $|\langle+|\frac{r_{kx}'r_{kz}'}{r_k'^5}|-\rangle_v|$ & $|\langle+|\frac{r_{ky}'r_{kz}'}{r_k'^5}|-\rangle_v|$ \\
\hline
\textit{Abinit} & 0.00 & 0.00 & 0.82 & 0.83 & 0.83 & 0.00 & 0.00 \\
\hline
\textit{RHF} & 0.00 & 0.00 & 0.51 & 0.51 & 0.51 & 0.00 & 0.00 \\
\hline
\end{tabular}
\caption{Intravalley and intervalley integrals for the S nucleus regime. The index and unit are the same as Table \ref{data1}.} \label{data2}
\end{table}

\end{widetext}

\subsection{Evaluation based on first principle calculated Bloch functions using Abinit}

We also numerically evaluated the integrals in the hyperfine interaction (\ref{HhfA}) using the Abinit all electron (AE) wave function. The results are also given in Tables \ref{data1} and \ref{data2}. In deriving the AE wave function, we choose a three dimensional (3D) unit cell.  The unit cell is like what we choose in Appendix \ref{intervalleycoupling}, but here all the lattice vectors are expended to 3D space, that is, $\vec{a}_1=(3.193,0,0), \vec{a}_2=(\frac{3.193}{2},\frac{3.193\sqrt3}{2},0)$, and $\vec{a}_3=(0,0,18.804)$. The Mo atom is located at $(\frac{3.193}{2},\frac{3.193\sqrt3}{6},1.567)$, while the two S atoms are at $(0,0,0)$ and $(0,0,3.134)$. The unit is $\mathring{A}$. Abinit gives us the periodic part of the Bloch states on $120\times120\times720$ discrete points which cover this 3D unit cell.

From Table \ref{data1} and \ref{data2}, we clearly see that the results from the numerical Abinit calculation agree well with the ones in Appendix B1 using the Roothaan-Hartree-Fock wavefunctions for the atomic orbitals.

\subsection{Corrections beyond the on-site contribution}

\begin{figure}[tbp]
\includegraphics[width=1.0\columnwidth]{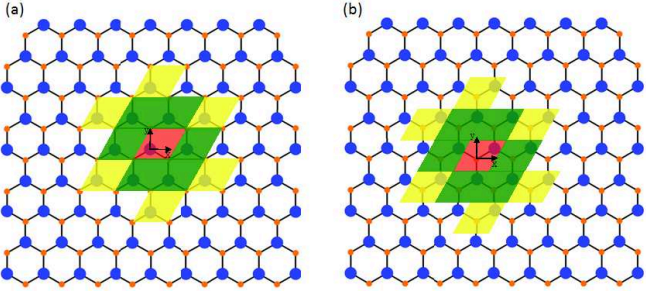}
\caption{Schematics of the on-site unit cell (red diamond), six nearest neighbors ones (green diamonds), and six next nearest neighbors ones (yellow diamonds) for the evaluation of the hyperfine interaction with the (a) transition metal and (b) chalcogen nuclear spin. Large blue balls denote metal sites and small orange balls denote chalcogen sites. The metal site in the red diamond of (a) corresponds to $r_k=0$ in Eq. (\ref{onsite}), and the chalcogen site in the red diamond of (b) corresponds to $r_k'=0$. In the evaluation of the on-site contribution, the range of integral $V$ in Eq. (\ref{onsite}) corresponds to the red diamond. In the evaluation of the nearest (next nearest) neighbor contribution, $V$ in Eq. (\ref{onsite}) corresponds to the sum of the green (yellow) diamonds.}
\label{neighborsFigure}
\end{figure}

Using the Abinit AE wave function, we examine here the corrections beyond the on-site contribution to the hyperfine interaction. Here we list some numerical results in calculating the integrals by involving more neighboring unit cells in Table \ref{shortrang}. One can see that for the integrals related to Mo nuclei the correction from all nearest neighbor unit cells (c.f. Fig. \ref{neighborsFigure}) is about $0.1\%$, and the next nearest neighbors' correction is even smaller. We check all non-vanishing integrals and find the correction is of the same order. Therefore, we conclude that the hyperfine interaction between electron and Mo nuclear spins are well counted within an on-site unit cell. For the integrals related to S nuclei, the nearest neighbors' correction can be $10\%$, and the next nearest neighbors' correction is about $1\%$. This does not affect very much the magnitude of the hyperfine interaction we estimated.
\begin{table}[!htb]
\begin{tabular}{|c|c|c|c|}
\hline
\hline
Integrals for Mo& $\langle+|\frac{r^2_{kz}}{r^5_k}|+\rangle_c$ & $\langle+|\frac{r^2_{kx}}{r^5_k}|+\rangle_c$ & $\langle+|\frac{r^2_{ky}}{r^5_k}|+\rangle_c$ \\
\hline
\textit{n.n.} & 0.0133 & 0.1070 & 0.1195 \\
\hline
\textit{n.n.n.} & 0.0007 & 0.0193 & 0.0185 \\
\hline
Integrals for S& $\langle+|\frac{r_{kz}'^2}{r_k'^5}|+\rangle_c$ & $\langle+|\frac{r_{kx}'^2}{r_k'^5}|+\rangle_c$ & $\langle+|\frac{r_{ky}'^2}{r_k'^5}|+\rangle_c$ \\
\hline
\textit{n.n.} & 0.0440 & 0.0599 & 0.0731 \\
\hline
\textit{n.n.n.} & 0.0034 & 0.0165 & 0.0158 \\
\hline
\end{tabular}
\caption{Integral corrections contributed from the six nearest neighbors (n.n.) unit cells and the six next nearest neighbors (n.n.n.) ones. The unit is $\mathring{A}^{-3}$ (the same as in Table \ref{data1} and \ref{data2}). See Fig. \ref{neighborsFigure} for the illustration of the n.n. and n.n.n. unit cells.} \label{shortrang}
\end{table}

\subsection{Analysis from the Rotational Symmetry}

In this following, we analyze the integrals involved the hyperfine Hamiltonian (\ref{HhfA}) based on the symmetry properties of the Bloch wave functions.
Under a $C_3$ rotation centered at the $k$-th M nucleus, we have
\begin{equation}
\begin{array}{rl}
\displaystyle C_3\Psi_{\tau\sigma}^{c}(\vec{r})=& \displaystyle \Psi_{\tau\sigma}^{c}(\vec{r}), \\
\displaystyle C_3\Psi_{\tau\sigma}^{v}(\vec{r})=& \displaystyle e^{i\tau\frac{2\pi}{3}}\Psi_{\tau\sigma}^{v}(\vec{r}).
\end{array}
\end{equation}

For the intravalley integrals in the conduction band,
\begin{eqnarray}
\left\langle+\right\vert\frac{r_{kx}^{2}}{r_{k}^{5}}\left\vert+\right\rangle_{c}&=&\int d\vec{r}\Psi_{+}^{c*}(\vec{r})\frac{r_{kx}^{2}}{r_{k}^{5}}\Psi_{+}^{c}(\vec{r}) \notag \\
&&=\int  d\vec{r}|C_3\Psi_{+}^{c}(\vec{r})|^{2}\frac{C_3(r_{kx}^{2})}{C_3(r_{k}^{5})} \notag \\
&&=\frac{1}{4}\left\langle+\right\vert\frac{r_{kx}^{2}}{r_{k}^{5}}\left\vert+\right\rangle_{c}+\frac{3}{4}\left\langle+\right\vert\frac{r_{ky}^{2}}{r_{k}^{5}}\left\vert+\right\rangle_{c} \notag \\
&&+\frac{\sqrt{3}}{2}\left\langle+\right\vert\frac{r_{kx}r_{ky}}{r_{k}^{5}}\left\vert+\right\rangle_{c}.
\end{eqnarray}
Similarly, we have
\begin{eqnarray}
\left\langle+\right\vert\frac{r_{kx}r_{ky}}{r_{k}^{5}}\left\vert+\right\rangle_{c}&=&-\frac{\sqrt{3}}{4}\left\langle+\right\vert\frac{r_{kx}^{2}}{r_{k}^{5}}\left\vert+\right\rangle_{c}+\frac{\sqrt{3}}{4}\left\langle+\right\vert\frac{r_{ky}^{2}}{r_{k}^{5}}\left\vert+\right\rangle_{c} \notag \\
&&-\frac{1}{2}\left\langle+\right\vert\frac{r_{kx}r_{ky}}{r_{k}^{5}}\left\vert+\right\rangle_{c}.
\end{eqnarray}
From the above two equations we find that
\begin{equation}
\begin{array}{l}
{\displaystyle \langle+|\frac{r_{kx}^{2}}{r_{k}^{5}}|+\rangle_{c}={\displaystyle \langle+|\frac{r_{ky}^{2}}{r_{k}^{5}}|+\rangle_{c}}},\\
{\displaystyle \langle+|\frac{r_{kx}r_{ky}}{r_{k}^{5}}|+\rangle_{c}={\displaystyle 0.}}
\end{array}
\end{equation}
Other integrals can be worked out in the same way. We find
that $\langle+|\frac{r_{kx}r_{kz}}{r_{k}^{5}}|+\rangle_{c}=\langle+|\frac{r_{ky}r_{kz}}{r_{k}^{5}}|+\rangle_{c}=0$.

The same relations hold for the intervalley integrals,
\begin{equation}
\begin{array}{l}
{\displaystyle \langle+|\frac{r_{kx}^{2}}{r_{k}^{5}}|-\rangle_{c}={\displaystyle \langle+|\frac{r_{ky}^{2}}{r_{k}^{5}}|-\rangle_{c}}},\\
\displaystyle \langle+|\frac{r_{kx}r_{ky}}{r_{k}^{5}}|-\rangle_{c}=\langle+|\frac{r_{kx}r_{kz}}{r_{k}^{5}}|-\rangle_{c}=\langle+|\frac{r_{ky}r_{kz}}{r_{k}^{5}}|-\rangle_{c}=0.
\end{array} \notag
\end{equation}

In the valence band subspace, the intravalley integrals
are similar to those in the conduction band subspace.
However, it is different for the intervalley integrals. We
find that
\begin{eqnarray}
\left\langle+\right\vert\frac{r_{kx}^{2}}{r_{k}^{5}}\left\vert-\right\rangle_{v}&=& -\left\langle+\right\vert\frac{r_{ky}^{2}}{r_{k}^{5}}\left\vert-\right\rangle_{v}, \notag \\
\left\langle+\right\vert\frac{r_{kx}r_{ky}}{r_{k}^{5}}\left\vert-\right\rangle_{v}&=&i\left\langle+\right\vert\frac{r_{ky}^{2}}{r_{k}^{5}}\left\vert-\right\rangle_{v}, \notag \\
\left\langle+\right\vert\frac{1}{r_{k}^{3}}\left\vert-\right\rangle_{v}&=&\left\langle+\right\vert\frac{r_{kz}^{2}}{r_{k}^{5}}\left\vert-\right\rangle_{v}=0, \notag \\
\left\langle+\right\vert\frac{r_{ky}r_{kz}}{r_{k}^{5}}\left\vert-\right\rangle_{v}&=& i\left\langle+\right\vert\frac{r_{kx}r_{kz}}{r_{k}^{5}}\left\vert-\right\rangle_{v}.
\end{eqnarray}

The Bloch wave functions under the $C_3$ rotation around X nucleus have the following relations,
\begin{equation}
\begin{array}{rl}
\displaystyle C'_3\Psi_{\nu\sigma}^{c}(\vec{r})=& \displaystyle e^{i\nu\frac{2\pi}{3}}\Psi_{\nu\sigma}^{c}(\vec{r}), \\
\displaystyle C'_3\Psi_{\nu\sigma}^{v}(\vec{r})=& \displaystyle e^{-i\nu\frac{2\pi}{3}}\Psi_{\nu\sigma}^{v}(\vec{r}). \\
\end{array}
\end{equation}
We find that the intravalley integrals have the same property as for the M nucleus. In the following we list the relations of intervalley integrals both for conduction and valence band subspaces,
\begin{eqnarray}
\left\langle+\right\vert\frac{1}{r_{k}'^{3}}\left\vert-\right\rangle_{c}&=&\left\langle+\right\vert\frac{r_{kz}'^{2}}{r_{k}'^{5}}\left\vert-\right\rangle_{c}=0, \notag \\
\left\langle+\right\vert\frac{r_{kx}'^{2}}{r_{k}'^{5}}\left\vert-\right\rangle_{c}&=&-\left\langle+\right\vert\frac{r_{ky}'^{2}}{r_{k}'^{5}}\left\vert-\right\rangle_{c}, \notag
\end{eqnarray}
\begin{eqnarray} \left\langle+\right\vert\frac{r_{kx}'r_{ky}'}{r_{k}'^{5}}\left\vert-\right\rangle_{c}&=&i\left\langle+\right\vert\frac{r_{ky}'^{2}}{r_{k}'^{5}}\left\vert-\right\rangle_{c}, \notag \\ \left\langle+\right\vert\frac{r_{ky}'r_{kz}'}{r_{k}'^{5}}\left\vert-\right\rangle_{c}&=&i\left\langle+\right\vert\frac{r_{kx}'r_{kz}'}{r_{k}'^{5}}\left\vert-\right\rangle_{c}, \notag
\end{eqnarray}
\begin{eqnarray}
\left\langle+\right\vert\frac{1}{r_{k}'^{3}}\left\vert-\right\rangle_{v}&=&\left\langle+\right\vert\frac{r_{kz}'^{2}}{r_{k}'^{5}}\left\vert-\right\rangle_{v}=0, \notag \\ \left\langle+\right\vert\frac{r_{kx}'^{2}}{r_{k}'^{5}}\left\vert-\right\rangle_{v}&=&-\left\langle+\right\vert\frac{r_{ky}'^{2}}{r_{k}'^{5}}\left\vert-\right\rangle_{v}, \notag
\end{eqnarray}
\begin{eqnarray} \left\langle+\right\vert\frac{r_{kx}'r_{ky}'}{r_{k}'^{5}}\left\vert-\right\rangle_{v}&=&-i\left\langle+\right\vert\frac{r_{ky}'^{2}}{r_{k}'^{5}}\left\vert-\right\rangle_{v},
\notag \\
\left\langle+\right\vert\frac{r_{ky}'r_{kz}'}{r_{k}'^{5}}\left\vert-\right\rangle_{v}&=&-i\left\langle+\right\vert\frac{r_{kx}'r_{kz}'}{r_{k}'^{5}}\left\vert-\right\rangle_{v}.
\end{eqnarray}

By comparing the relations obtained from symmetry analysis with the numerical estimation in Table \ref{data1} and \ref{data2}, one can find that they agree very well. We note that the relative errors between the two numerical estimations becomes larger when we deal with the S nuclei regime, which are possibly due to the small magnitude of the integrals. In the numerical results obtained from Abinit wave function, we find a finite intervalley interaction in the valence band subspace for Mo nuclei, which is predicted to be 0 in Roothaan-Hartree-Fock method. However, the values are so small that they are probably resulted from the calculation errors.

\subsection{Optimized atomic orbital functions} \label{orbital}
In the RHF method, we have the radial part of Mo-5s, Mo-4d, and S-3p orbitals as follows, \cite{Clementi1974table,McLean1981table}
\begin{widetext}
\begin{equation}
\begin{array}{rl}
\displaystyle f_{5s}(r)=& \displaystyle 7.22802 e^{-42.7425 r} + 116.248 e^{-36.2098 r} r -110.945 e^{-19.9247 r} r + 8.81419 e^{-52.4536 r} r^2 +21.0326 e^{-14.652 r} r^2 + \\
& \displaystyle 104.34 e^{-8.1505 r} r^2 -3214.89 e^{-23.3229 r} r^3 - 37.4997 e^{-5.1157 r} r^3 -4.14133 e^{-3.4917 r} r^3 + 316.287 e^{-13.6857 r} r^4 +\\
& \displaystyle 0.749325 e^{-2.3571 r} r^4 + 0.120996 e^{-1.4897 r} r^4 + 0.006006 e^{-0.9661 r} r^4,\\ [2ex]
\displaystyle f_{4d}(r)=& \displaystyle 81.5662 e^{-22.9005 r} r^2 + 329.352 e^{-12.658 r} r^2 +40.5766 e^{-6.0525 r} r^2 - 14.1228 e^{-3.5536 r} r^2 +227.235 e^{-9.7486 r} r^3 - \\
& \displaystyle 4.0599 e^{-2.7024 r} r^3 -0.451674 e^{-1.7351 r} r^3 - 0.0220095 e^{-1.1346 r} r^3,\\ [2ex]
\displaystyle f_{3p}(r)=& \displaystyle 3.55739 e^{-22.6414 r} r - 19.0356 e^{-10.4197 r} r -9.64606 e^{-6.116 r} r - 7.56414 e^{-4.4156 r} r +45.9701 e^{-17.3448 r} r^2 + \\
& \displaystyle 4.35629 e^{-2.6496 r} r^2 +1.39527 e^{-1.6975 r} r^2 + 0.179256 e^{-1.1477 r} r^2.\\ [2ex]
\end{array}
\end{equation}
\end{widetext}
These atomic orbitals are used to form the Bloch states. Note that in the above expressions $r$'s are all in the atomic unit here. We need a transform of the unit in order to calculate the integrals in the hyperfine interaction.

\end{document}